\documentclass[aps,prb,twocolumn,superscriptaddress,10pt]{revtex4-1}

\usepackage{graphicx}
\usepackage{amsmath}
\usepackage{amssymb}
\usepackage{amsfonts}
\usepackage{bm}
\usepackage{psfrag}
\usepackage{pstricks}

\usepackage{color}

\pdfoutput=1

\begin{document}
\title{Hybrid functional study of non-linear elasticity and internal strain in zincblende III-V materials}

\author{Daniel S.~P.~Tanner}
\email{daniel.tanner@tyndall.ie}
\affiliation{Tyndall National Institute, Lee Maltings, Dyke Parade, Cork T12 R5CP, Ireland}

\author{Miguel~A. Caro}
\affiliation{Department of Electrical Engineering and Automation, Aalto University, Espoo 02150, Finland} 
\affiliation{Department of Applied Physics, Aalto University, Espoo 02150, Finland}

\author{Stefan Schulz}
\affiliation{Tyndall National Institute, Lee Maltings, Dyke Parade, Cork T12 R5CP, Ireland}

\author{Eoin P.~O'Reilly}
\affiliation{Tyndall National Institute, Lee Maltings, Dyke Parade, Cork T12 R5CP, Ireland}
\affiliation{Department of Physics, University College Cork, Cork T12 YN60, Ireland}

\begin{abstract}
We investigate the elastic properties of selected zincblende III-V semiconductors. 
Using hybrid functional density functional theory we calculate the second and third
order elastic constants, and first and second-order internal strain tensor components
for Ga, In and Al containing III-V compounds. For many of these parameters, there are 
no available experimental measurements, and this work is the first to predict their
values. The stricter convergence criteria for the calculation of higher order elastic 
constants are demonstrated, and arguments are made based on this for extracting these 
constants via the calculated stresses, rather than the energies, in the context of 
plane-wave-based calculations. 
The calculated elastic properties are used to determine the strain regime at which higher
order elasticity becomes important by comparing the stresses predicted by a lower and a
higher order elasticity theory.
Finally, the results are compared with available experimental 
literature data and previous theory.

\end{abstract}

\date{\today}


\pacs{78.67.De, 73.22.Dj, 73.21.Fg, 77.65.Ly, 73.20.Fz, 71.35.-y}

\maketitle

\section{Introduction}
Elastic constants are fundamental material parameters, a knowledge of which is essential for
the design and understanding of semiconductor materials and devices. For example, the electronic and optical properties
of semiconductor heterostructures are strongly influenced by the strain state of their active regions.\cite{Reilly89} This
strain state depends on the lattice mismatch between the constituent compounds, and on the relative
magnitude of their elastic constants.\cite{Nye_Book} Elastic constants are also 
necessary for: the determination of the material composition of heterostructures by X-ray diffraction;\cite{MoVi09}
assessing the critical thickness and strain relaxation in devices;\cite{HoCo07} modeling the behaviour of dislocations;\cite{Kittel_8th}
the characterisation of piezoelectric resonators;\cite{Mems}
and the parameterisation of interatomic potentials\cite{Keating66} used for the calculation of strain fields in supercells
containing millions of atoms. 

For crystals which lack inversion symmetry, standard macroscopic elasticity theory does not fully describe
the position of their atoms under strain, and \emph{internal strain}\cite{Born_dynamic,Klein62,CousinsInner} occurs. 
Internal strain is a displacement between sublattices in a crystal. It is described, for a 
particular material, by the components of the internal strain tensor. Knowledge of these material
parameters is essential for any semi-emprical atomistic modelling which requires the equilibrium atomic positions of
strained structures, and has, for instance, particular importance for the 
piezoelectricity of a crystal.\cite{Birman58,Martin_Piezo}

For many device and material applications, infinitesimal strain theory,\cite{LANDAU1986} in which there appear only second-order elastic
constants (SOEC) and first-order internal strains, is sufficient to describe the elastic properties. 
This means that the crystal energy can be accurately expressed to second-order in the strain, with the SOECs as coefficients,
and that the internal strain can be described accurately to first-order in the strain, 
with the first-order internal strain tensor components (ISTCs) as coefficients.
However, as the strain in the system increases, its energy (internal strain) can no longer be accurately described
using only a second (first) order expansion in the strain, and higher order terms need to be accounted for.
The lowest order of such corrections are third-order elastic constants (TOEC) for the macroscopic strain energy, and second-order ISTCs
for the internal strain. The strain magnitudes at which these corrections to the energy and internal relaxation 
become important will depend on the relative magnitudes of the higher and lower order coefficients of the strains. 

Recent studies have shown that third-order contributions to the
elastic energy are necessary to correctly model the relaxation, strain state,
and thus optical propeties, of several technologically important heterostructures 
such as InGaAs/GaAs,\cite{PrKi98,FrDo00,ElFa02,MaWa04,Lepkowski2008} InGaN/GaN, and 
GaN/AlGaN.\cite{LoMa07,Lepkowski2004,LeMa05,Lepkowski2008,ShAg98} 
Similar effects can be expected in other highly lattice-mismatched 
nanostructured systems, such as the InSb/GaSb quantum dot (QD) system.\cite{DeTa07} 
Furthermore, second-order ISTCs have been shown to play
an important role in the piezoelectric response in many materials.\cite{CaSc15,MiPo06}
Other phenomena related to lattice anharmonicity, 
such as: phonon-phonon or electron-phonon interactions;\cite{HikiRev} 
thermal expansion;\cite{HikiRev} stress or temperature dependent elastic 
response;\cite{HikiRev} and pressure dependence of optic mode frequencies,\cite{CousinsThesis} 
also require the use of TOECs and second-order ISTCs.

Because of this wide application, there has long been interest in the 
measurement or theoretical determination of TOECs.\cite{Brug64,KeatingThird,Birch47, Wallace1970} 
Typically, TOECs are measured using the velocity of sound waves through a crystal under uniaxial 
or hydrostatic stress,\cite{ThBr64} analysed via the finite strain theory of 
Murnaghan.\cite{Murnaghan_Book,MurnaghanPaper} 
However, there are often very large uncertainties in such measurements,\cite{JoDu06} and 
these difficulties are compounded for brittle or metastable semiconductor materials,\cite{CousinsThesis,CoGe91,LaGu11}
which are often of interest for device applications. Similarly, for the \emph{internal} strain, 
of the materials addressed in this paper, only GaAs and InSb have available experimental values of their single first
order ISTC (Kleinman parameter), due to the high precision of experimental observations 
required\cite{CousinsThesis,ChMo96} for the extraction of ISTCs;
there are no experimental evaluations of any second-order ISTCs.
Thus, given the experimental difficulty in the measurement of ISTCs in general,
and the importance of non-linear elasticity for 
the accurate description of the electronic and optical properties of 
technologically relevant semiconductors and their connected heterostructures, there is strong 
motivation for the theoretical determination of TOECs and first and second-order ISTCs. 

Early calculations of TOECs involved the use of pseudopotential\cite{SuTe68} and 
interatomic force potential methods;\cite{KeatingThird} however,
since the work of Nielsen and Martin,\cite{NiMa85} first principles calculations 
have become an increasingly popular route towards the calculation of TOECs. In 
recent years, \emph{ab-initio} methods have been used to determine the TOECs for ultrahard
materials such as diamond;\cite{HmWi16,Niel86} materials of which it is 
difficult to obtain high quality single crystals, like the metastable cubic-phase 
nitride materials;\cite{LoMa07} and technologically important materials for which
experimental TOEC measurements are sparse, such as InAs and GaAs.\cite{SoSc98,NiMa85,LoMa07,Lepkowski2008}

In this work, we present first-principles calculations of SOECs, TOECs, as 
well as first and second-order ISTCs,
of a range of III-V zincblende (ZB) semiconductor compounds. 
The calculations are carried out using density functional theory (DFT) within the 
Heyd-Scuseria-Ernzerhof (HSE) hybrid-functional approach.\cite{HeSc03} 
We demonstrate that higher order elastic properties 
require a higher resolution of calculation parameters for their accurate evaluation, and extend arguments 
present in the literature for the use of the stress method for
the extraction of elastic constants to the case of TOECs.
Moreover, we show the importance of third-order effects in the strain regimes 
relevant to a sample InSb/GaSb heterostructure system, and demonstrate for
other materials the errors incurred by the use of a linear theory. Finally, our 
results are compared, where possible, with previous experimental and theoretical results. 
Overall, we find very good agreement with previously reported 
experimental and theoretical literature data.

The paper is organised as follows: in Section~\ref{sec:theory} we present the finite strain theory in which the 
TOEC and second-order ISTCs are defined; in Section~\ref{sec:Comp} we discuss our computational framework, giving the
specifics of the DFT implementation and discuss the different nature of convergence of TOECs compared with SOECs,  
including a comparison of constants extracted from the stress-strain approach 
with those extracted via the total energy;
in Section~\ref{sec:Results} we present the calculated SOECs, TOECs and first and second-order ISTCs, 
make comparisons with recent experiment and theory, and apply the
extracted TOECs to address the question as to the strain 
regime in which non-linear elasticity need be used; finally, in Section~\ref{sec:Summary}, we summarise and conclude.

\section{Overview of finite strain theory}
\label{sec:theory}
In this section we review the aspects of finite strain theory necessary for the calculation and 
discussion of TOECs and second-order ISTCs. In Sec.~\ref{sec:theory_elas}, we apply finite strain
to the discussion of the macroscopic elasticity of crystals, and in Sec.~\ref{sec:theory_inner},
we outline the theory describing the internal strain resulting from a given applied finite strain. 

\subsection{Elasticity}
\label{sec:theory_elas}
In  solid state physics, the description of third-order elasticity is conventionally
achieved via the Lagrangian strain formalism.\cite{Brug64,NiMa85,LoMa07,HmWi16}
The application of Lagrangian stresses and strains to the theory of elasticity 
with finite deformations has been developed by Murnaghan,\cite{MurnaghanPaper,Murnaghan_Book}
and applied to cubic crystals by Birch.\cite{Birch47}

The deformation gradient tensor, $F$, marks the starting point of all strain formalisms.
It describes the deformation of a material, including rotations, when the coordinates
of that system are transformed. If the position of a point in a material is given by
$\mathbf{a}$, and after strain is at the position $\mathbf{x}$, then the deformation tensor may be defined as:\cite{Nye_Book}
\begin{equation} \label{eq:J}
 F = F_{ij} = \frac{\partial x_{i}}{\partial a_{j}} .
\end{equation}
This relates simply to the linear strain tensor as:
\begin{equation} 
 F_{ij} = \varepsilon_{ij}+\delta_{ij} ,
\end{equation}
where $\delta_{ij}$ is the Kronecker delta, and $\varepsilon$ is the small, or infinitesimal, strain tensor.
While this simple relation between the infinitesimal strain and the deformation is very useful and attractive,
the conceptual underpinning of the infinitesimal strain (that it measures the relative changes of lengths in the material)
becomes increasingly invalid with increasing strain. Thus, in the regime of larger strains, where third-order elasticity
becomes relevant, Lagrangian strains are employed. The Lagrangian strain tensor, $\eta_{ij}$, is related
to the deformation by:\cite{Birch47}
\begin{equation}
 \eta_{ij}=\frac{1}{2}\left(F_{ip}F_{jp}-\delta_{ij}\right) ,
\end{equation}
where Einstein summation notation is used.
In cases where the infinitesimal strain tensor is known, the following useful matrix
relation may be used to determine the Lagrangian strain tensor:\cite{Birch47}
\begin{equation} \label{eq:eps_to_eta}
 \mathbf{\eta} = \mathbf{\varepsilon}+\frac{1}{2}\mathbf{\varepsilon}^{2} .
\end{equation}
The TOECs are conventionally defined in terms of the expansion of the free energy density
in these Lagrangian strains. For a cubic crystal, this energy density is given by:\cite{Brug64,Birch47,LoMa07}
\begin{multline} \label{eq:cubic_ThirdOrder}
 \rho_{0}E = \frac{1}{2}C_{11}\left(\eta_{1}^{2}+\eta_{2}^{2}
 +\eta_{3}^{2}\right)
 + \frac{1}{2}C_{44}\left(\eta_{4}^{2}+\eta_{5}^{2}+\eta_{6}^{2}\right) \\
 + C_{12}\left(\eta_{1}\eta_{2}+\eta_{1}\eta_{3}+\eta_{2}\eta_{3}\right) 
 + \frac{1}{6}C_{111}\left(\eta_{1}^{3}+\eta_{2}^{3}+\eta_{3}^{3}\right) \\
 + \frac{1}{2}C_{112}\left(\eta_{2}\eta_{1}^{2}+\eta_{3}\eta_{1}^{2}+\eta_{2}^{2}\eta_{1}+\eta_{3}^{2}\eta_{1}+\eta_{2}\eta_{3}^{2}+\eta_{2}^{2}\eta_{3}\right)\\ 
 + C_{123}\eta_{1}\eta_{2}\eta_{3} + \frac{1}{2}C_{144}\left(\eta_{1}\eta_{4}^{2}+\eta_{2}\eta_{5}^{2}+\eta_{3}\eta_{6}^{2}\right) \\
 + \frac{1}{2}C_{155}(\eta_{2}\eta_{4}^{2} +\eta_{3}\eta_{4}^{2}+\eta_{1}\eta_{5}^{2}+\eta_{3}\eta_{5}^{2}+\eta_{1}\eta_{6}^2+\eta_{2}\eta_{6}^{2}) \\
 + C_{456}\eta_{4}\eta_{5}\eta_{6} .
\end{multline}
Here $\rho_{0}$ is the mass density of the unstrained material, $E$ is the Helmholtz free energy per unit mass, 
and the various $C_{ij}$ and $C_{ijk}$ are the second and third-order isentropic elastic constants, respectively.
We have also above employed Voigt~\cite{Voigt,Nye_Book} notation, which, using the symmetry of the strain tensor,
makes the convenient contraction of indices: 11$\rightarrow$1, 22$\rightarrow$2, 33$\rightarrow$3, 32$\rightarrow$4, 
13$\rightarrow$5, 12$\rightarrow$6. The derivatives of this energy density, $\rho_{0}E$,  with respect to 
the $\eta_{i}$ provide equations relating the Lagrangian stresses, $t_{i}$, to the Lagrangian strains, 
$\eta_{i}$, via the elastic constants:
\begin{equation} \label{eq:Lag_StressStrain}
 t_{i} = \rho_{0}\frac{\partial E}{\partial \eta_{i}} .
\end{equation}
Thus, the general expressions for the Voigt components of the Lagrangian stress in terms of an arbitrary
Lagrangian strain on a cubic crystal are:
\begin{equation} \label{eq:t_eta}
 \begin{split}
 t_{1}& = C_{11}\eta_{1}+C_{12}\left(\eta_{2}+\eta_{3}\right)+\frac{1}{2}C_{111}\eta_{1}^{2}
 +\frac{1}{2}C_{112}(2\eta_{2}\eta_{1}\\
 &+2\eta_{3}\eta_{1} +\eta_{2}^{2}+\eta_{3}^{2}) +C_{123}\eta_{2}\eta_{3}+
 \frac{1}{2}C_{144}\eta_{4}^{2} \\
 &+\frac{1}{2}C_{155}\left(\eta_{5}^{2}+\eta_{6}^{2}\right), \\
 t_{4} &= C_{44}\eta_{4}+C_{144}\eta_{1}\eta_{4}+C_{155}\left(\eta_{2}\eta_{4}+\eta_{3}\eta_{4}\right)+C_{456}\eta_{5}\eta_{6} , 
 \end{split}
\end{equation}
with $t_{2,3}$ and $t_{5,6}$, given by cyclic permutations of the indices of $t_{1}$ and $t_{4}$, respectively.

However, when the stresses on a strained supercell are calculated via DFT 
using the Hellmann-Feynman theorem\cite{NiMa83} or from an interatomic potential calculation, it is the stresses 
on the faces of the deformed cell that are
obtained; these are the Cauchy stresses, $\sigma$. Therefore, in order to use eqs.(\ref{eq:t_eta}) to extract
elastic constants from DFT data, the Lagrangian stress must be related to the Cauchy stress:\cite{Murnaghan_Book}
\begin{equation} \label{eq:sigma}
 t = det\left(F\right)F^{-1}\sigma\left(F^{T}\right)^{-1} .
\end{equation}

Hence, by either measuring the energy or stress of a cubic crystal as a function of applied Lagrangian strain, eqs.~(\ref{eq:cubic_ThirdOrder})
,(\ref{eq:t_eta}) and (\ref{eq:sigma}) may be used to obtain values for the elastic constants.

Having established the finite strain formalism required for the discussion of TOECs, in the next section
we describe the application of this formalism to the description of non-linear inner elasticity in cubic crystals.

\subsection{Internal strain}
\label{sec:theory_inner}
Non-linear internal strain involves a second, rather than first-order description
of the internal strain in terms of the Lagrangian strain.
To achieve this description of sublattice displacement to second-order in the regime
of large strains for ZB semiconductors, we will use the formalism introduced by Cousins.\cite{CousinsInner,CousinsThesis}

Taking the ZB primitive cell, and letting the atom at the origin remain fixed, the 
position of the second atom, after strain, is given by:
\begin{equation} \label{eq:Transformation}
 \mathbf{r} = F\mathbf{r}_{0} + \mathbf{u} ,
\end{equation}
where $\mathbf{r}_{0}$ is its equilibrium position, and $\mathbf{u}$
represents the internal strain vector.
Although this transformation completely specifies the deformed positions geometrically, 
the $\mathbf{u}$ are not suitable parameters in which to expand the scalar energy. 
This is because they lack rotational invariance. Given that the internal strain represents 
the atomic configuration which minimises the energy of the ZB crystal under shear strain, 
a rotationally invariant description of the internal strain is needed. This is obtained 
through use of what Cousins~\cite{CousinsInner} calls the \emph{inner displacement}. This is given by:
\begin{equation}
 \boldsymbol{\xi} = F^{T}\mathbf{u} .
\end{equation}
Because this inner displacement occurs in response to internal forces arising from 
the application of finite strain, each inner displacement can be expressed as a 
Taylor series in the components of the strain:
\begin{equation} \label{eq:zetai}
 \xi_{i} = A_{iJ}\eta_{J} + \frac{1}{2}A_{iJK}\eta_{J}\eta_{K} .
\end{equation}
Here, Voigt notation has been employed for the elements of the finite strain tensor, and 
the Einstein summation convention is again utilised. 
The subscripts relating to the strain are denoted by capitals, whilst those 
relating to the Cartesian coordinate of the inner displacement are denoted by the lower-case $i$.
The $A_{iJ}$ and $A_{iJK}$ are the first and second-order \emph{internal strain tensors}, respectively. 
Cousins~\cite{CousinsThesis,CousinsSymmetry} gives the
form of these tensors for a ZB crystal. The first-order 
internal strain tensor may be expressed conveniently in matrix notation:
\begin{equation} \label{eq:AiJ}
 A_{iJ} = \begin{pmatrix}
          0 & 0 & 0 & A_{14} & 0      & 0 \\
          0 & 0 & 0 & 0      & A_{14} & 0 \\
          0 & 0 & 0 & 0      & 0      & A_{14} 
          \end{pmatrix} .
\end{equation}
We note that for small strains, $F\approx I$, the identity matrix, and $\boldsymbol{\xi}\approx\mathbf{u}$,
where $u_{i} = -\frac{a_{0}}{2}\zeta\varepsilon_{jk}$ ($u_{1} = -\frac{a_{0}}{2}\zeta\varepsilon_{23}$, 
$u_{2} = -\frac{a_{0}}{2}\zeta\varepsilon_{13}$, $u_{3} = -\frac{a_{0}}{2}\zeta\varepsilon_{12}$), and 
thus $A_{14}=-\frac{a_{0}}{4}\zeta$, where $\zeta$ is the well known Kleinman parameter.\cite{Klein62}

A matrix representation is not possible for $A_{iJK}$, but there are only three independent non-zero components
which are:\cite{CousinsSymmetry}
\begin{eqnarray} \label{eq:AiJK}
 & A_{114} = A_{225} = A_{336}, \\
 & A_{156} = A_{246} = A_{345}, \\
 & A_{124} = A_{235} = A_{316} = A_{134} = A_{215} = A_{326} . \label{eq:last}
\end{eqnarray}
Substituting eqs.~(\ref{eq:AiJ})-(\ref{eq:last}) into eq.~(\ref{eq:zetai}) 
yields an expression for the value of $\boldsymbol{\xi}$ which
minimises the strain energy of a ZB crystal for a given applied finite strain:
\begin{equation} \label{eq:internal_Zeta}
\begin{split}
 \boldsymbol{\xi} &= A_{14}\begin{pmatrix} \eta_{4} \\ \eta_{5} \\ \eta_{6} \end{pmatrix} + \frac{A_{114}}{2}\begin{pmatrix} \eta_{1}\eta_{4} \\ \eta_{2}\eta_{5} \\ \eta_{3}\eta_{6} \end{pmatrix} \\
 & +\frac{ A_{124}}{2}\begin{pmatrix} \eta_{4}\left(\eta_{2}+\eta_{3}\right) \\ \eta_{5}\left(\eta_{3}+\eta_{1}\right) \\ \eta_{6}\left(\eta_{1}+\eta_{2}\right) \end{pmatrix}
 + \frac{A_{156}}{2}\begin{pmatrix} \eta_{5}\eta_{6} \\ \eta_{4}\eta_{6} \\ \eta_{4}\eta_{5} \end{pmatrix} .
 \end{split}
\end{equation}

If, for a given primitive ZB unit cell, the atomic positions corresponding to the 
energetic minimum for a particular Lagrangian strain branch are known, 
eq.~(\ref{eq:internal_Zeta}) can be used in fitting procedures to obtain the first 
and second-order ISTCs.



In the next section, the manner in which the presented finite strain theory is 
applied to deformed unit cells is described. In addition, the details of the DFT
caculations performed to obtain the stresses on, and energies of, these deformed unit
cells are presented, along with a discussion of the different calculational criteria
needed for the accurate calculation of elastic constants and internal strain tensor components.

\section{Computational Method}  \label{sec:Comp}
In this section we discuss the computational method used to calculate the
SOECs, TOECs and first and second-order ISTCs.

First, in Sec.~\ref{Def}, the deformations applied to each ZB unit cell are presented.
This is accompanied by a description of the strains and stresses associated with these deformations
via the finite strain theory introduced in the previous section. In Sec.~\ref{Calc}, the 
details of the DFT calculations are given. This is followed in Sec.~\ref{sec:Conv} by an analysis of the 
convergence of the calculations with respect to k-point grid density, plane-wave cutoff energy, lattice constant,
and applied strain range. In particular, it is demonstrated and explained that a higher resolution of calculation in terms
of k-point grid density, cutoff energy, and lattice constant, is needed to achieve convergence of TOECs when
compared with that needed for the calculation of SOECs. Whether to calculate the
elastic constants via the calculated stresses or energies is discussed. Our results 
show that, using the energy method, the convergence of TOECs is much slower with 
respect to cutoff energy, k-points and applied strain range when compared to that 
exhibited by the stress strain method. Consequently, the energy method is significantly 
more computationally expensive. A similar behaviour has been reported in the literature 
for SOECs,\cite{Miguel_Stress} where it was identified that the slower convergence of the 
energy method occurs due to changes in the \textit{k}-point basis set used, as the cut-off energy is kept fixed in the calculations,
while the unit cell size/shape is being varied. By contrast,the stresses, when 
calculated according to the Hellmann-Feynmann theorem,
are computed implicitly for a fixed basis set (since they 
are computed at fixed lattice vectors). We then find, 
given the already high computational demands, that the 
effect of the changing basis set in the energy calculations 
is strongly enhanced when calculating TOECs, as further discussed in Section~\ref{Kpoint_energy}.


\subsection{Applied deformations} \label{Def}
For the extraction of all SOECs, TOECs, and first and second-order ISTCs, data from the following
applied strain branches~\cite{CaSc15} were used:
\begin{equation} \label{eq:Strain_Branches}
 \begin{aligned}
  \boldsymbol{\epsilon}^{(1)} &\equiv \left(0,0,0,\beta,\beta,\beta\right), \\
  \boldsymbol{\epsilon}^{(2)} &\equiv \left(\alpha,0,0,\beta,0,0\right), \\
  \boldsymbol{\epsilon}^{(3)} &\equiv \left(0,\alpha,0,\beta,0,0\right), \\
  \boldsymbol{\epsilon}^{(4)} &\equiv \left(0,\alpha,\alpha,\beta,0,0\right), \\
  \boldsymbol{\epsilon}^{(5)} &\equiv \left(\alpha,\alpha,\alpha,\beta,\beta,\beta\right) .
 \end{aligned}
\end{equation}
The corresponding deformations of the unit cell were chosen such that, according to the symmetries of 
eqs.~(\ref{eq:AiJ})-(\ref{eq:internal_Zeta}), at least one independent determination of
each of the second-order ISTCs would be obtained from the resultant inner
displacements. Deformations chosen in this way also enabled several pathways to the determination
of each of the SOECs and TOECs from the stresses on the unit cells.
For the calculation of all elastic constants and ISTCs, $\alpha$ and $\beta$ are varied
in the following manner: In the strain branch $\boldsymbol{\epsilon}^{(1)}$, $\beta$ is 
varied in steps of 0.01 from $-$0.04 to 0.04, resulting in a total of 9 strain points. 
For each of the remaining branches, $\alpha$ ($\beta$) is varied from $-$0.02 ($-$0.04) to 0.02 (0.04) in steps 
of 0.01 (0.02), comprising a total of 25 points. Each value of $\alpha$ and $\beta$ 
is associated with six stress components, a total energy
value, and the position vectors of the atoms in the ZB primitive cell. 
To ascertain the form of the relation between the deformation parameters $\alpha$ and $\beta$, and the
stress, energy, and relaxed atomic positions, the Lagrangian strains corresponding to each deformation
branch must be determined. Using the tensor relation, eq.~(\ref{eq:eps_to_eta}), the Lagrangian strains, $\boldsymbol{\eta}^{(i)}$, obtained
from the strain branches, $\boldsymbol{\epsilon}^{(i)}$, are given by:
\begin{equation} \label{eq:Lagrainge_Branches}
 \begin{aligned}
  \boldsymbol{\eta}^{(1)} &\equiv \left(\frac{\beta^{2}}{4},\frac{\beta^{2}}{4},\frac{\beta^{2}}{4},\beta+\frac{\beta^{2}}{4},\beta+\frac{\beta^{2}}{4},\beta+\frac{\beta^{2}}{4}\right), \\
  \boldsymbol{\eta}^{(2)} &\equiv \left(\alpha+\frac{\alpha^{2}}{2}, \frac{\beta^{2}}{8} , \frac{\beta^{2}}{8}, \beta, 0, 0\right), \\
  \boldsymbol{\eta}^{(3)} &\equiv \left(0,\alpha^{\prime}, \frac{\beta^{2}}{8} , \beta+\frac{\alpha\beta}{2},0,0\right), \\
  \boldsymbol{\eta}^{(4)} &\equiv \left(0,\alpha^{\prime},\alpha^{\prime},\beta+\alpha\beta,0,0\right), \\
  \boldsymbol{\eta}^{(5)} &\equiv \left(\alpha^{\prime}+\frac{\beta^{2}}{8}, \alpha^{\prime}+\frac{\beta^{2}}{8}, \alpha^{\prime}+\frac{\beta^{2}}{8}, \beta^{\prime},\beta^{\prime},\beta^{\prime}\right),
 \end{aligned}
\end{equation}
where the notation $\alpha^{\prime} = \alpha+\frac{\alpha^{2}}{2}+\frac{\beta^{2}}{8}$ and $\beta^{\prime} = \beta+\frac{\beta^{2}}{4}+\alpha\beta$, has been used
for compactness. 

With these five strain branches there are: five energy equations from eq.~(\ref{eq:cubic_ThirdOrder}), 
from which all SOEC and TOECs may be obtained; fourteen unique stress component equations, from 
which all SOEC and TOEC may be obtained in multiple ways using eqs.~(\ref{eq:t_eta}); and five
equations for inner displacements, from eq.~(\ref{eq:internal_Zeta}). 
We do not present these twenty four long equations here in the interest of brevity. 
However, to illustrate the methodology, we will present in the results section a sample subset of these equations,
truncated to second order in the deformation parameters. A more detailed description of the full set of stress and inner displacement
fitting equations is given in Ref.~[\onlinecite{MyThesis}].

The first three of the Lagrangian strain branches of eq.(~\ref{eq:Lagrainge_Branches}) 
allow, via the relaxed atomic positions, determination of $A_{14}$ 
(the Kleinman parameter within the finite strain formalism, $\approx-\frac{\zeta a_{0}}{4}$), 
and all three of the second-order ISTCs. Furthermore, because $\alpha$ and $\beta$ are varied 
independently, these first three branches also furnish, via the different 
components of the stress tensor, multiple independent determinations of each of the 
SOECs, and all but one of the TOECs. The remaining TOEC, $C_{123}$, can then be determined from the stress in the $x$ direction
associated with $\boldsymbol{\eta}^{(4)}$.
The stresses and relaxed atomic positions of the more complicated strain branch, $\boldsymbol{\eta}^{(5)}$, 
are not used to obtain any new values for the elastic constants, but to check 
the overall accuracy and consistency of the full set of elastic constants or ISTCs 
by substituting in particular values derived from the other branches, and performing
fits for the remaining constants. Since the fitting will be dependent on the accuracy 
of the substituted constants, the agreement of the result with previous 
fittings indicates accurate determination of all those substituted constants, as well 
as those newly obtained.

\begin{figure*}[ht]
   \centering
   \includegraphics[width=0.9\textwidth]{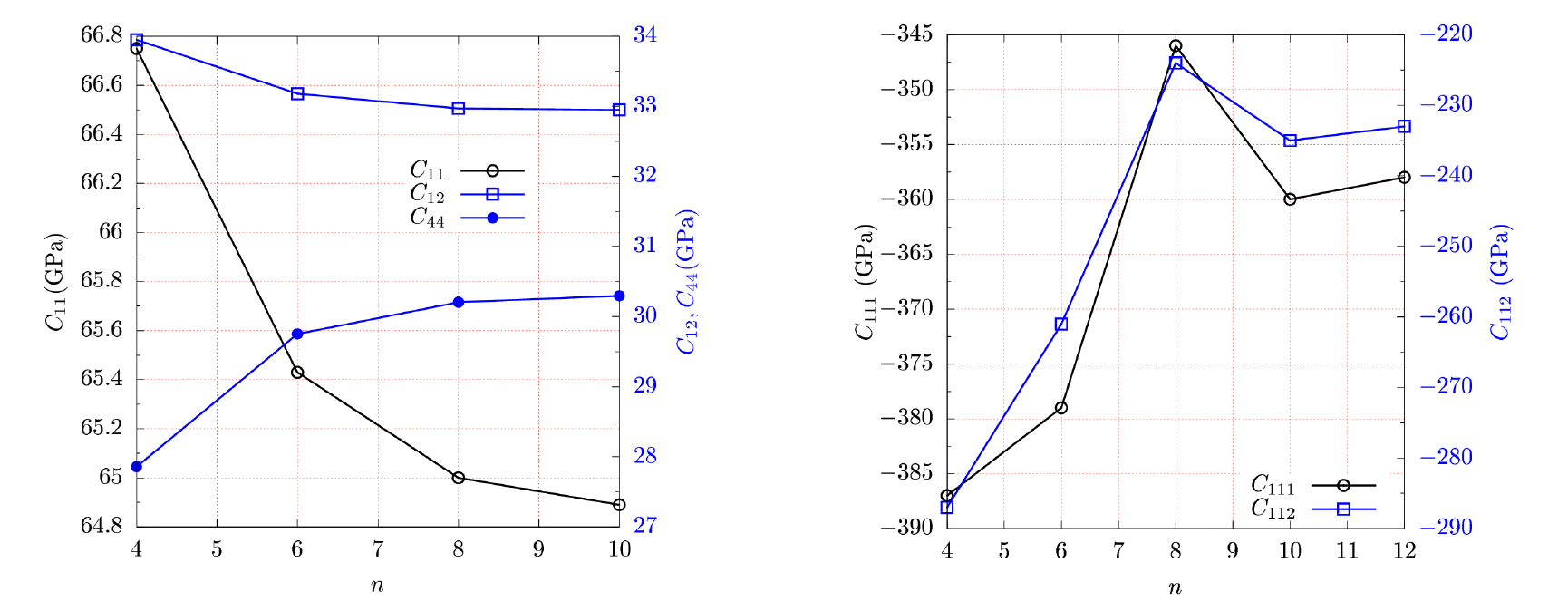}
   \caption{Comparison of convergence with k-point density of \emph{stress-extracted} 
   InSb SOECs (left) and TOECs (right). The calculations were performed at a cutoff energy of 600 eV,
   on a k-point grid of $n\times n\times n$, with $C_{11}$,$C_{12}$,$C_{111}$ and $C_{112}$ all determined from an applied strain 
   of $\boldsymbol{\epsilon}=(\alpha,0,0,0,0,0)$, with $\alpha$ varied between $\pm$2\% in steps of 1\% 
   and $C_{44}$ is determined via a shear strain $\boldsymbol{\epsilon}=(0,0,0,\beta,\beta,\beta)$, 
   with $\beta$ varied between $\pm$4\% in steps of 2\%.}
   \label{fig:InSb_StressConv}
\end{figure*}

\subsection{DFT framework for the calculation of energy, stress, and relaxed atomic positions} \label{Calc}
To obtain the energies, stresses, and relaxed atomic positions associated with each of the above strain branches,
DFT calculations were performed on the (deformed) ZB unit cells using the 
Heyd-Scuseria-Ernzerhof (HSE) hybrid-functional approach.\cite{HeSc03} The 
calculations were carried out using the software package $\mathrm{VASP}$.\cite{KrFu96} 
A screening parameter, $\mu$, of 0.2 \AA$^{-1}$, and an exact exchange mixing parameter, $\alpha$, 
of 0.25, were utilised; these correspond to $\mathrm{VASP}$'s HSE06 version of the HSE functional.
More details are given in Ref.[\onlinecite{CaSc15}].

For the calculation of material parameters, DFT within the HSE scheme 
offers improved accuracy over standard Kohn-Sham approaches to the exchange energy.\cite{HePa2011}
For instance, it circumvents the well known band gap problem of
LDA and generalised gradient approximation (GGA) implementations. Moreover, 
HSE-DFT has been shown to give improved predictions of elastic and lattice properties 
of solids over LDA and GGA implementations.\cite{PaMa06,RaMo15} 
Fitting to the DFT data using eqs.~(\ref{eq:cubic_ThirdOrder}), (\ref{eq:t_eta}) and 
(\ref{eq:internal_Zeta}), yields values for the SOECs and TOECs 
as well as first and second-order ISTCs as described above.

For the determination of elastic constants, we choose to fit to the stress-strain equations (eq.~(\ref{eq:t_eta}))
rather than the energy-strain equations (eq.~(\ref{eq:cubic_ThirdOrder})) for reasons 
of greater accuracy and efficiency,\cite{Miguel_Stress,NiMa83,NiMa85, HmWi16} which we
will demonstrate below. In general, the stress method is suited to efficient 
calculation of the elastic constant tensor because a single DFT calculation 
yields the full stress tensor, with its six unique components, and six equations 
to fit to. However, a single DFT calculation produces only one scalar energy, 
with one equation available to fit to.
Thus the elastic constants can be obtained via the stress 
method efficiently from one calculation, whereas from the energy method, several separate calculations 
are needed for the same number of constants. Furthermore, in terms of accuracy, 
the equations relating these elastic constants to the strains will be a lower order 
polynomial in the strain, and therefore easier to fit when dealing with very small strains. 
Finally, as will be corroborated in the next section: in a plane wave-based DFT implementation
using a fixed cutoff energy in its plane-wave expansion at different k-points and different lattice vectors, 
the number of k-points and cutoff energy needed in a given calculation to 
obtain converged values of the elastic constants is lower for those 
calulated via the stress method than for those calculated via the total energy method.\cite{Miguel_Stress}
Consequently, elastic constants can be determined to a desired accuracy at less computational 
expense using the stress method. Moreover, these issues of convergence are even more
pronounced for TOECs and second-order ISTCs than SOECs and first-order ISTCs,
as will be demonstrated below.


\subsection{Convergence of Results} \label{sec:Conv}
In this section, we analyse the convergence of our results, and show that the 
aforementioned advantages in convergence which the stress method exhibits over 
the energy method, demonstrated in the literature in the context of SOECs,\cite{Miguel_Stress,NiMa83,NiMa85} 
are even more dramatic for the TOECs. First, in Sec~\ref{Kpoint_energy}, we show that 
TOECs require a higher resolution of calculation than SOECs:
that elastic constants extracted via the stress method converge faster with respect to k-point mesh density
and cutoff energy than those extracted via the energy method; and that the elastic
constants presented here, calculated using the stress method, are converged (where 
those extracted via the energy method, from the same calculation, may not be). These convergence 
tests are shown using InSb as a model system, which of the studied materials is the slowest 
to converge with k-point density and cutoff energy. Thus, the presented results for InSb validate 
also the convergence of the other III-V materials studied here. Second, in Sec.~\ref{strain_presh} the 
dependence of the extracted elastic constants on the equilibrium pressures and applied strain 
ranges is investigated. The increased sensitivity of TOECs to these calculation parameters when 
compared with SOECs is highlighted, and this increased sensitivity is shown to be worsened 
when elastic constants are extracted through use of the energy method. 
Finally, the suitability of our choice of applied strain range and allowed equilibrium pressure 
are confirmed. 

\begin{figure*}[ht]
   \centering 
   \includegraphics[width=0.9\textwidth]{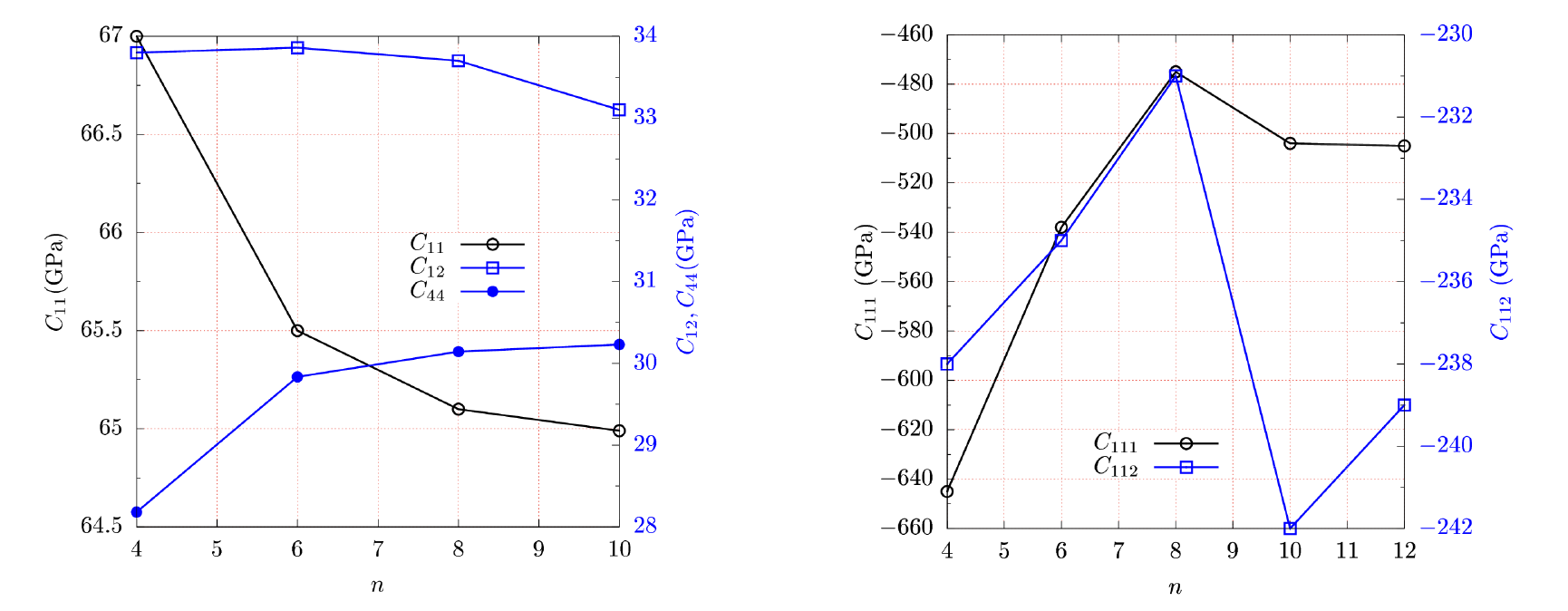}
   \caption{Convergence with k-point density of InSb (left) SOECs and (right) TOECs \emph{
   extracted using the energy method}. The calculations were performed at a cutoff energy of 600 eV, 
   on an $n\times n\times n$
   k-point grid. $C_{11}$,$C_{12}$,$C_{111}$ and $C_{112}$ are all determined from 
   an applied strain of $\boldsymbol{\epsilon}=(\alpha,0,0,0,0,0)$, 
   with $\alpha$ varied between $\pm$2\% in steps of 1\% 
   and $C_{44}$ is determined via a shear strain $\boldsymbol{\epsilon}=(0,0,0,\beta,\beta,\beta)$, 
   with $\beta$ varied between $\pm$4\% in steps of 2\%.}
   \label{fig:InSb_EnergyConv}
\end{figure*} 

\subsubsection{Convergence with k-points and cutoff energy}\label{Kpoint_energy}
Compared to the effects of linear elasticity, third-order 
elasticity gives rise to smaller changes in stress, energy, and atomic positions. 
Therefore, it can be expected that convergence of TOECs will require denser k-point 
grids and higher cutoff energies than is the case for SOECs. We find that
this is indeed the case, as illustrated in Fig.~\ref{fig:InSb_StressConv}, where, 
for InSb, the slower convergence of the stress-extracted TOECs with respect to 
k-point density may be immediately inferred from the different scales. Here the cutoff energy is
fixed at 600 eV. On closer inspection of Fig.~\ref{fig:InSb_StressConv},
one finds that the percentage changes in $C_{11}$ and $C_{12}$ on going from a
$6\times6\times6$ to an $8\times8\times8$ k-point grid are both 1\%, whilst for $C_{111}$ and $C_{112}$ 
the values change by 10\% and 17\%, respectively. Therefore, while
an $8\times8\times8$ k-point grid may be sufficient to obtain converged SOECs, the calculation
of TOECs requires a higher k-point density.
Examining the percentage change in the calculated 
constants when going from an $8\times8\times8$ to a $10\times10\times10$ k-point mesh, convergence
of both SOECs and TOECs is apparent. For the SOECs a negligible difference of ~0.5\% exists, 
whilst for the TOECs, $C_{111}$ and $C_{112}$, the values differ by only 
4\% and 5\%, respectively. To further corroborate convergence of the TOECs,
we note the negligible change on increasing the k-point density from $10\times10\times10$ to $12\times12\times12$;
the differences being $<$1\% for both TOECs.
With these small changes between subsequent grid sizes, 
we conclude that a grid of $10\times10\times10$ is 
sufficient to converge the stress extracted elastic 
constants, at a cutoff energy of 600~eV.

While Fig.~\ref{fig:InSb_StressConv} establishes convergence of the elastic constants
extracted through the stress method, Fig.~\ref{fig:InSb_EnergyConv} justifies the choice
of extracting the elastic constants using the stress rather than energy by showing the poorer
convergence of the energy method. 
Figure~\ref{fig:InSb_EnergyConv} shows that, using the energy method, the TOECs also converge
much slower when compared with the SOECs. Also similarly to the convergence of the stress-extracted
constants, the energy-extracted TOECs are also clearly converged with respect to k point density by a 
$10\times10\times10$ k-point grid density. However, in this case, the TOECs exhibit much larger fluctuations
at lower grid densities. For example, at a grid of $4\times4\times4$, the energy extracted $C_{111}$ is 
28\% lower than the converged value, whilst the stress-extracted $4\times4\times4$ $C_{111}$ is only 8\% lower than its
converged value.
In addition to the fact that the energy values are converging slowly, we note,
more importantly, that they are also unconverged at this cutoff energy, with final $C_{111}$ and 
$C_{112}$ values of -504 and -242~GPa, respectively, compared to the converged stress-extracted values of -360 and -235~GPa.
\begin{table}[hb] 
\centering  
\begin{tabular}{ccccc}
\hline\hline
$E_{cut}~(eV) $    & $C^{t}_{11}~(GPa)$ & $C^{E}_{11}~(GPa)$ & $C^{t}_{111}~(GPa)$   & $C^{E}_{111}~(GPa)$ \\ \hline
400                & 65.2$\pm$0.1       & 65   $\pm$1        & $-$379 $\pm$11        & $-$1108 $\pm$15     \\ 
600                & 64.89$\pm$0.03     & 65.0 $\pm$0.2      & $-$360 $\pm$4         & $-$504  $\pm$34     \\ 
1000               & 64.8 $\pm$0.1      & 64.9 $\pm$0.1      & $-$359 $\pm$6         & $-$445  $\pm$21     \\ 
\hline\hline
\end{tabular}
\caption{Effect of cutoff energy on elastic constants, $C_{11}$ and $C_{111}$, of InSb calculated using the stress and energy method.
A superscript of $t$ denotes the stress method, and $E$ denotes the energy method. Calculations all performed with a
k-point grid of $10\times10\times10$. } \label{tab:Cutoff Energy Convergence}
\end{table}

\begin{figure*}[ht]
\centering
\includegraphics[width=0.9\textwidth]{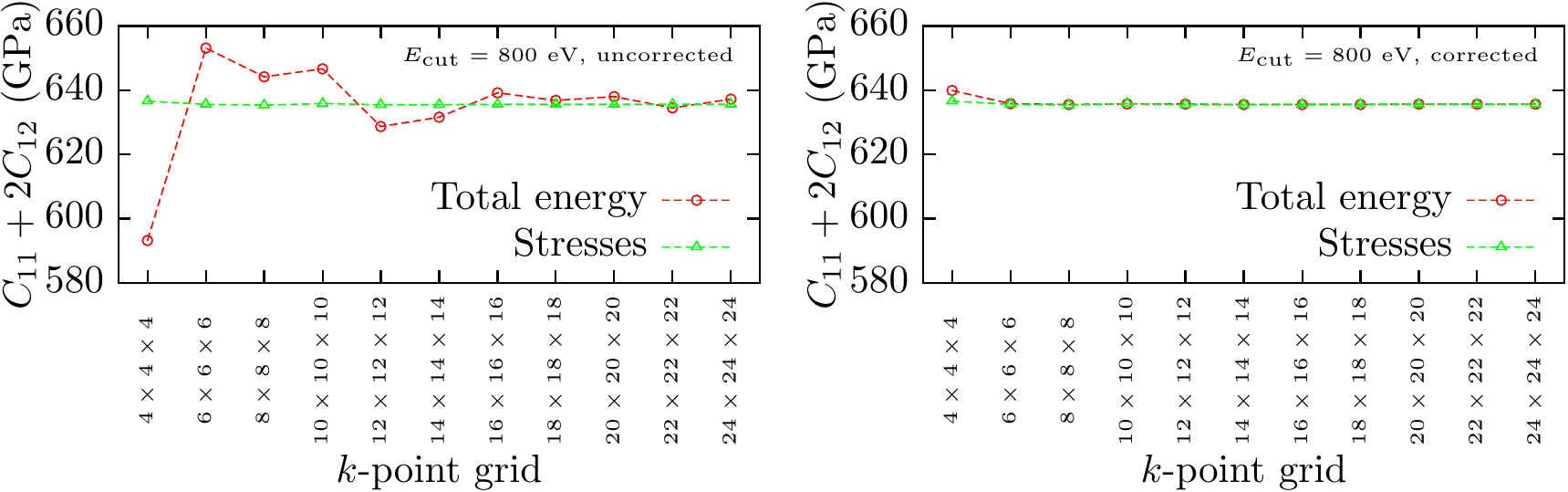}
\caption{LDA calculation of the bulk modulus ($C_{11}+2C_{12}$) of AlN, using both the stress and energy method. 
(left) Results for a fixed cutoff energy of 800~eV, 
and (right) with a strain-corrected cutoff around 800 eV, for the same
system (hydrostatically strained ZB AlN).}
\label{047}
\end{figure*}

Therefore, in a second step, we analyse the impact of the cutoff energy on the elastic constants.
Table~\ref{tab:Cutoff Energy Convergence} shows the effect of increasing the cutoff 
energy, with a fixed k-point grid of $10\times10\times10$, on the 
calculated $C_{11}$ and $C_{111}$ values. The superscripts, $t$, and $E$, 
refer to constants extracted via the stress and energy methods, respectively. 
The numbers following the ``$\pm$'' are the fitting errors.
The table shows that for both the energy and stress method, a cutoff of 400 eV is more than sufficient to obtain
converged SOECs. However, for the TOEC $C_{111}$, only the stress extracted $C_{111}$ 
is converged. As with Figs.~\ref{fig:InSb_StressConv} and \ref{fig:InSb_EnergyConv}, 
the table shows that: TOECs generally require higher cutoff energies than SOECs; energy extracted TOECs
require higher cutoff energies for a given accuracy than stress extracted TOECs;  
that a cutoff energy of 600 eV is sufficent to obtain converged TOECs (for InSb) using 
the stress method; and that even for a cutoff energy of 1000 eV, the 
energy method still does not yield a converged value for $C^{E}_{111}$, as can be seen 
by comparison with $C^{t}_{111}$.

The slower convergence of the energy extracted parameters with respect to those extracted via the stresses
is due to the larger impact of the changing plane-wave basis set on the strain energy than on the stress.\cite{Miguel_Stress}
The total energy results can in principle
be corrected by using a (in general anisotropic) strain-dependent cutoff
energy. For small strains, $\pm 1\%$ change in lattice vectors corresponds
to $\sim \mp 2\%$ change in cutoff energy. This is because, for a cutoff energy
$E_{cut}$, only those plane waves that obey the condition $\arrowvert\mathbf{G}+\mathbf{k}\arrowvert < G_{cut}$, where
$\mathbf{G}$ is a reciprocal lattice translation, are included in the basis, with:
\begin{align}
E_\text{cut} = \frac{\hbar^2}{2 m} {G_\text{cut}}^2.
\label{046}
\end{align}
Modifying the cut-off energy to maintain a fixed basis set leads to a remarkably
improved agreement between total energy and stress methods, as shown in
Fig.~\ref{047}, where, to illustrate this point we have performed LDA calculations
of the bulk modulus of AlN using energy and stress, with and without the cutoff energy
correction.

Unfortunately, this cut-off energy correction can only be easily implemented for
hydrostatic strain (i.e. only allows to calculate bulk modulus)
because this is the only case where the basis set changes isotropically, 
consistent with eq.~\ref{047}.
Because it avoids these issues, the stress method should
therefore be used for reliable, consistent, and computationally
inexpensive calculation of elastic constants.

Finally, we note that InSb, being the heaviest and 
softest material, will require the highest resolution of calculation 
in terms of cutoff energy and k-point mesh. Thus, the convergence indicated 
in Fig.~\ref{fig:InSb_StressConv} and Table~\ref{tab:Cutoff Energy Convergence} 
also serves to confirm that the chosen cutoff energy and k-point 
density are appropriate for the other materials.

\subsubsection{Convergence with applied strain range and equilibrium pressure}\label{strain_presh}

In addition to their sensitivity to k-point grid density and cutoff energy,
TOECs also exhibit a more pronounced dependence on the residual pressure 
at the assumed equilibrium lattice constant, and the range of strain applied 
to the system in order to calculate them. Because the 
demonstration of this point requires a large number of calculations, in this section 
we analyse the TOECs and SOECs of AlN. For this compound of lighter atoms, the stress 
and total energy calculations are computationally less expensive than for InSb.
Unless stated otherwise, all convergence tests are performed at a cutoff energy 
of 600 eV and on a $10\times10\times10$ k-point mesh.  

\begin{table}[b] 
\centering  
\begin{tabular}{ccccc}
\hline\hline
 $a_{0}$ (\AA) &  $P_{0}$(kB)    & $C_{11}$(GPa)  & $C_{111}$(GPa) & $C_{111}^{\prime}$(GPa) \\ \hline
  4.3643       &   0.5927    & 310$\pm$3        & -1471$\pm$169    &  -1122$\pm$3        \\  
  4.3646       &   0.1521    & 309.56$\pm$0.40  & -1212$\pm$43     &  -1123$\pm$3        \\ 
  4.3647       &   0.0051    & 309.49$\pm$0.03  & -1125$\pm$3      &  -1122$\pm$3        \\ 
  4.3648       &  $-$0.1369  & 309.41$\pm$0.37  & -1037$\pm$40     &  -1118$\pm$3        \\ 
  4.3651       &  $-$0.5752  & 309$\pm$2        & -778$\pm$164     &  -1117$\pm$4        \\ 
\hline\hline
\end{tabular}
\caption{Effect of residual pressure due to insufficient lattice relaxation on $C_{111}$ and $C_{11}$ of AlN. 
The accompanying errors are the least squares fitting errors. $C_{111}^{\prime}$ is the value for $C_{111}$ extracted
using a fitting which accounts for the equilibrium pressure. The calculations were performed with a cutoff 
energy of 600~eV and a k-point grid density of $10\times10\times10$.} \label{tab:ResidualPressure}
\end{table}
The issue of lattice constant relaxation is examined in Table~\ref{tab:ResidualPressure}, 
where the elastic constants $C_{11}$ and $C_{111}$, extracted
using the stress method, are shown for different equilibrium lattice constants and pressures.
The $C_{111}$ and $C_{11}$ values displayed are the result of applying 
the strain branch $\boldsymbol{\varepsilon} = (\alpha,0,0,0,0,0)$ 
with $\alpha$ varied from $-0.02$ to $0.02$ in steps of $0.01$.
The pressure denoted by $P_{0}$ in the table is the calculated residual pressure 
on the ZB primitive cell with the lattice 
constant given in the first column. 
The values preceded by the ``$\pm$'' are least squares fitting errors.

Table~\ref{tab:ResidualPressure} shows that, when optimising the lattice 
constant by minimising the absolute value of the pressure on the unit cell, 
the magnitude of the pressure below  which we  may                                             
accurately extract elastic constants from the stress, using standard fitting methods, 
is lower for TOECs than for SOECs. 
In columns three and four of Table~\ref{tab:ResidualPressure} are presented the results
of fitting the equation $t_{1}=(\frac{C_{111}}{2}+\frac{C_{11}}{2})\alpha^{2}+C_{11}\alpha$ directly
to the DFT data, which include this 'equilibrium' pressure, $P_{0}$, as the pressure corresponding
to a strain of 0\%.
Here the variation of AlN's $C_{111}$ value with residual ``equilibrium'' pressure
may be contrasted with the constancy of the corresponding $C_{11}$ value. 
As the residual pressure increases, so does the value of $C_{111}$ deviate from the value
obtained at lowest pressure, along with increasing fitting errors. For a lattice constant
change of -0.0004~\AA~ from the lowest pressure lattice constant, a 30\% error is incurred in
$C_{111}$.
This may be attributed to related factors such as: the small magnitudes 
of the contribution of the TOEC to the total stress (at $\alpha=0.02$, $\frac{C_{111}}{2}\alpha^{2}$=2.25~kB),
of which the initial pressure is a significant fraction; and the tendency of higher order
polynomials to be more sensitive to noise in fitting. 

The errors so-incurred can be reduced by two means. The first is to modify the fitting equation
to account for the equilibrium pressure; i.e. by fitting  using the equation: 
$t_{1}=(\frac{C_{111}^{\prime}}{2}+ \frac{C_{11}}{2})\alpha^{2}+C_{11}\alpha+P_{0}$. The improvements induced
by this adjustment are evident in the stability with intial pressure of $C_{111}^{\prime}$ in column 
five of Table~\ref{tab:ResidualPressure}. The second way to reduce these errors is to ensure
the lattice has been relaxed to a sufficiently low pressure; while the origin adjustement
shown before more than solves the problem for AlN over the given pressure range, for softer 
materials, this sensitivity to initial pressure will be even more pronounced, with a given 
pressure corresponding to a higher strain, and this adjustement is less effective.
We thus impose more stringent criteria on the maximum pressures below which we 
consider a crystal to be relaxed, aiming for pressures below 0.1~kB, a fifth of the cutoff value of $\sim$0.5~kB
typically used for SOECs.\cite{ElaStic}

Another important calculation parameter to which the TOECs are sensitive is the range of strain 
applied to the unit cell.\cite{HmWi16,ElaStic} Applying strains over a larger range will 
produce larger changes in stress and energy from which the contribution of the
third-order terms will be more easily discerned; however, as the strain range is increased,
even higher order terms may begin to have an effect. Furthermore,
having a large strain range with a constant strain point density will require
a larger number of calculations. Thus, the strain range applied will need to be large enough
that the effect of the TOECs can be observed, but not so large that further higher order terms come into
play, or the calculation is prohibitively expensive; i.e., the optimal strain range is the minimum
strain range at which the effects of TOECs are appreciable.

\begin{table}[t] 
\centering  
\begin{tabular}{ccccc}
\hline\hline
$\alpha_{\textrm{max}}$  & $C^{t}_{11}$      & $C^{t}_{111}$  & $C^{E}_{111}$    & $C^{E}_{111}$~(1000~eV)\\ \hline
$\pm$2\%                 & 309.49 $\pm$0.03  & $-$1125 $\pm$3 & $-$2959 $\pm$381 &   $-$1223 $\pm$22 \\ 
$\pm$4\%                 & 310.0 $\pm$0.1    & $-$1125 $\pm$6 & $-$1627 $\pm$80  &   $-$1153 $\pm$6  \\
$\pm$6\%                 & 310.8 $\pm$0.2    & $-$1126 $\pm$8 & $-$1359 $\pm$31  &   $-$1137 $\pm$5  \\
$\pm$8\%                 & 311.9 $\pm$0.3    & $-$1128 $\pm$9 & $-$1261 $\pm$17  &   $-$1132 $\pm$5  \\ 
\hline\hline
\end{tabular}
\caption{Impact of range of applied strain in fitting data for AlN on SOEC 
$C^{t}_{11}$ extracted from the stresses, and TOECs $C^{t}_{111}$
and $C^{E}_{111}$ extracted from the stresses and energy, respectively. All 
elastic constants are in units of GPa. $C^{t}_{11}$, $C^{t}_{111}$ and $C^{E}_{111}$  
are calculated on a k-point grid of density $10\times10\times10$, and with a cutoff 
energy of 600 eV. $C^{E}_{111}$ (1000~eV) has the same settings but with the cutoff 
energy increased to 1000~eV.
} \label{tab:MaxStrain}
\end{table}

Table~\ref{tab:MaxStrain} shows the influence of the range of applied strain on the calculated
elastic constants of AlN. The superscript $t$ in this table refers to a stress 
extracted constant, and $E$ denotes an energy extracted constant. $\alpha_{\textrm{max}}$ 
denotes the maximum value of $\alpha$ in the applied strain of $\boldsymbol{\varepsilon}=(\alpha,0,0,0,0,0)$, 
with the data set comprising strains in increments of 1\% between $\pm\alpha_{max}$.
The stability of the stress extracted $C_{11}^{t}$ and $C_{111}^{t}$ in Table~\ref{tab:MaxStrain} reveals that the 
range of $\pm2$\% is large enough to yield measurable non-linearities in the stress, 
but not so large that higher order terms interfere with the fitting. The increasing 
influence of these unwanted higher order terms can be 
observed in the increasing errors of $C^{t}_{111}$. The rightmost two columns of the table show again
the shortcomings of the energy method when compared with the stress method for the extraction of 
TOECs; the energy extracted constants requiring larger strain ranges to lower the 
fitting error. The rightmost column shows the interrelation between the cutoff energy 
and the strain range. Small errors in the calculated free energy can have significant 
impact on the determined TOECs at small strain; increasing the cutoff energy 
reduces the scale of these errors, while increasing the strain range reduces 
their relative input in the total calculated change in energy. Overall, we see 
that the convergence of the TOEC values is clearly slower using the energy method.

Having justified our choice of the stress method over the energy method for the extraction
of SOECs and TOECs, and shown that our stress extracted constants are indeed converged, 
we present in the next section  the full set of SOECs and TOECs for all considered materials,
and discuss the results.

\section{Results}\label{sec:Results}
In this section the calculated SOECs, TOECs and first and second-order ISTCs are presented
and discussed.
First, in Section \ref{Res:elas}, the calculated elastic constants are presented, along with
plots validating the fittings used to obtain them. The extracted values are compared with 
previous experimental and theoretical literature results, and an analysis of the strains at which third-order
effects become important is made. In Section \ref{Res:Inner},
the calculated first and second ISTCs are reported.

\subsection{Elastic constants} \label{Res:elas}

Given below in eqs.~(\ref{eq:t_AlphaBeta}) and (\ref{eq:t_AlphaBeta_Shear}), 
are six sample stress fitting equations, truncated to second-order 
in the strain, each furnishing an independent determination of a subset of 
the nine independent elastic constants of a ZB crystal. Eqs.~(\ref{eq:t_AlphaBeta})
show three axial stress equations which may be used to determine simultaneously the SOECs: $C_{11}$ and $C_{12}$; and the TOECs: $C_{111}$, $C_{112}$
and $C_{123}$.
\begin{equation} \label{eq:t_AlphaBeta}
 \begin{aligned}
 t_{1}^{(2)}(\alpha,0) &= \frac{1}{2}\left(C_{11} + C_{111}\right)\alpha^{2}+ C_{11}\alpha,\\
 t_{2}^{(2)}(\alpha,0) &=\frac{1}{2}\left(C_{12} + C_{112}\right)\alpha^{2}+ C_{12}\alpha,\\
 t_{1}^{(4)}(\alpha,0) &= \left(C_{12} + C_{112} + C_{123}\right)\alpha^{2} + 2C_{12}\alpha  ,\\
 \end{aligned}
\end{equation}
and eqs.~(\ref{eq:t_AlphaBeta_Shear}) show three shear stresses 
which yield simultaneously values of the SOEC $C_{44}$, and TOECs $C_{144}$, $C_{155}$ and $C_{456}$:
\begin{equation} \label{eq:t_AlphaBeta_Shear}
 \begin{aligned}
 t_{4}^{(2)}(\alpha,\beta) &= C_{144}\alpha\beta+ C_{44}\beta , \\
 t_{4}^{(3)}(\alpha,\beta) &= \left(\frac{1}{2}C_{44} + C_{155}\right)\alpha\beta + C_{44}\beta , \\
 t_{4}^{(1)}(\beta) &=  \left(\frac{1}{4}C_{44}+C_{456}\right)\beta^{2} +C_{44}\beta .
 \end{aligned}
\end{equation}
Here the subscripts on the $t_{i}^{(n)}$ refer to the stress tensor component in Voigt
notation, and the superscripts, $n$, refer back to the strain branches, 
$\boldsymbol{\eta}^{(n)}$, in eq.~(\ref{eq:Lagrainge_Branches}). The zeros in brackets in eq.~(\ref{eq:t_AlphaBeta_Shear})
indicates that $\beta$ is set to 0, and only $\alpha$ is varied.

\begin{figure}[t]
\centering\includegraphics[width=0.85\columnwidth]{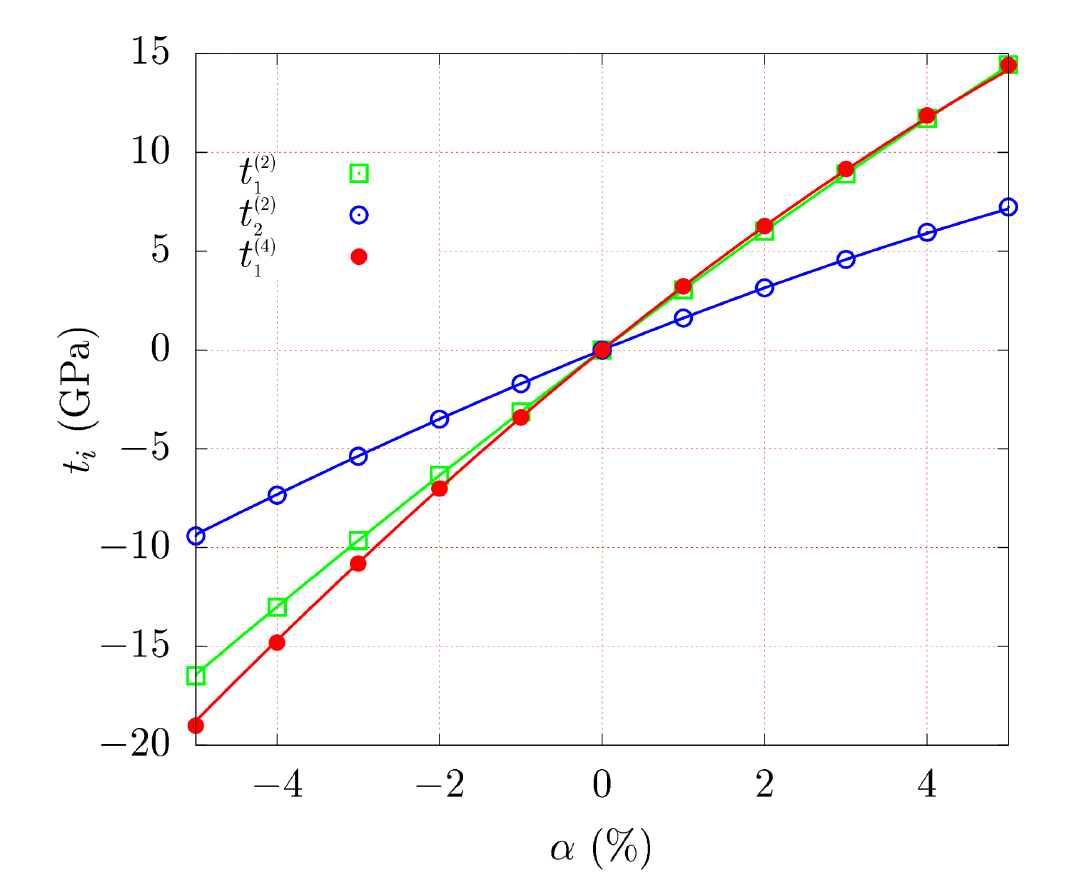}
\caption{Fitting Lagrangian stresses of eqs.~(\ref{eq:t_AlphaBeta}) to AlN HSE-DFT data. 
The points represent calculated DFT data, and the lines 
depict stresses determined from eqs.~(\ref{eq:t_AlphaBeta}). The used elastic constants
defining the functions in eqs.~(\ref{eq:t_AlphaBeta}) are obtained by fitting 
to the data out to $\pm2\%$, and are  
presented in Table~\ref{tab:Elastic_Consts}. }
\label{fig:Axial_Fitting}
\end{figure}

\begin{figure}[t]
\centering\includegraphics[width=0.85\columnwidth]{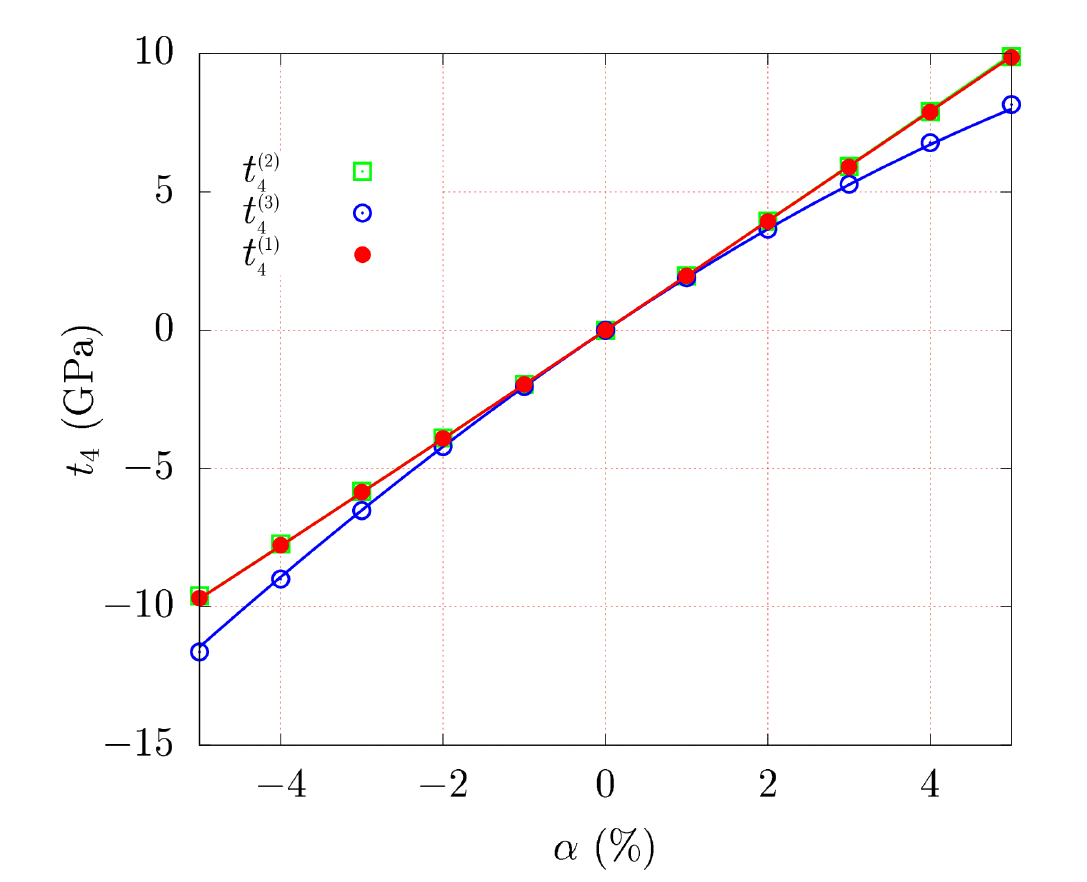}
\caption{Fitting shear stresses of eqs.(\ref{eq:t_AlphaBeta_Shear}) to one dimensional 
line scans of AlN HSE-DFT data. Here the data plotted are on the line $\alpha=\beta$. 
The points represent calculated DFT data points, and the lines represent stresses 
calculated via eqs.(\ref{eq:t_AlphaBeta_Shear}), with coefficients fitted from data sets
with $-2\%<\alpha<2\%$ and $-4\%<\beta<4\%$, using elastic constants 
presented in Table~\ref{tab:Elastic_Consts}.}
\label{fig:Coupled_Fitting}
\end{figure}

The solid lines in Figs.\ref{fig:Axial_Fitting} and \ref{fig:Coupled_Fitting} show the stresses on AlN unit cells as a 
function of strain, calculated using the expressions in eqs.~(\ref{eq:t_AlphaBeta}) and (\ref{eq:t_AlphaBeta_Shear}).
The figures also display the  stresses calculated by DFT for each strain as symbols. 
Note that the fitting of the coefficients in eqns.~(\ref{eq:t_AlphaBeta}) and 
(\ref{eq:t_AlphaBeta_Shear}) is not done on the data sets shown in the figure. 
These coefficients were obtained by two-dimensional fittings
to unsimplified untruncated stress equations using only data points in the range: $-2\%\leq\alpha\leq2\%$ and $-4\%\leq\beta\leq4\%$. 
The lines shown in the figure show then predictions for higher strain values. As the figures confirm,
not only is the fit very good at $\pm2\%$, but the line also matches the DFT data
points very well at higher strains.
The influence of non-linear effects may be inferred from 
the slight curvature and asymmetry of the lines.

By performing 
fittings to several stress relations,
the full set of SOECs and TOECs for all considered materials were obtained.
For $C_{11}$ and $C_{111}$, there are two independent determinations, from $t_{1}^{(2)}$ and $t_{2}^{(3)}$, and the 
values given in Tables~\ref{tab:SOEC} and \ref{tab:Elastic_Consts} are the averages of these two. 
The constants $C_{12}$ and $C_{112}$ have 
three independent determinations, $t_{2,3}^{(2)}$, $t_{1}^{(3)}$ and the values given in the table are the
averages of these. $C_{123}$ is obtained from the single fitting to $t_{1}^{(4)}$. For $C_{155}$, we extracted 
six separate values from the different stresses on the unit cells; the value given is the average of all these
very closely agreeing values. $C_{144}$ is given as an average over the values 
obtained from the three stresses $t_{1}^{(2)}$, $t_{4}^{(2)}$, and $t_{1}^{(3)}$. 
Finally, for all materials, $C_{44}$ and $C_{456}$ were obtained from $t_{4}^{(1)}$.

\begin{table*}[t] 
\centering  
\begin{tabular}{c|cccccc}
\hline\hline
     &               & $a_{0}$(\AA{})                             & $C_{11}$(GPa)                        & $C_{12}$(GPa)                         & $C_{44}$(GPa)                       & $\zeta$             \\ \hline
     & Prev theory   & 4.342\cite{WrNe95}                         & 304\cite{Wright97}, 282\cite{LoMa07} & 160\cite{Wright97}, 149\cite{LoMa07}  & 193\cite{Wright97}, 179\cite{LoMa07}& 0.55\cite{Wright97} \\ 
AlN  & Experimental  & 4.373\cite{As2010}, 4.38\cite{PeMo92}      & -                                    & -                                     & -                                   & -                   \\ 
     & Present       & 4.3647                                     & 309.47                               & 166.06                                & 196.90                              & 0.5385              \\ \hline\hline
     & Prev theory   & 5.417\cite{WaYe02}, 5.4735\cite{SinghBook} & 132.5\cite{WaYe03}                   & 66.7\cite{WaYe03}                     & 62.7\cite{WaYe03}                   & 0.604\cite{WaYe03}\\ 
AlP  & Experimental  & 5.46\cite{BeKo82}, 5.4635\cite{MadelHB}    & -                                    & -                                     & -                                   & -                 \\ 
     & Present       & 5.4713                                     & 138.25                               & 67.73                                 & 66.52                               & 0.5759            \\ \hline\hline
     & Prev theory   & 5.614\cite{WaYe02}                         & 113.1\cite{WaYe03}                   & 55.5\cite{WaYe03}                    & 54.7\cite{WaYe03}                    & 0.592\cite{WaYe03} \\ 
AlAs & Experimental  & 5.66139\cite{MadelHB}                      & 119.9\cite{MadelHB}                  & 57.5\cite{MadelHB}                   & 56.6\cite{MadelHB}                   & -                  \\ 
     & Present       & 5.6865                                     & 116.64                               & 55.62                                & 56.96                                & 0.5746             \\ \hline\hline
     & Prev theory   & 6.090\cite{WaYe02}                         & 85.5\cite{WaYe03}                    & 41.4\cite{WaYe03}                    & 39.9\cite{WaYe03}                    & 0.601\cite{WaYe03} \\ 
AlSb & Experimental  & 6.1355\cite{MadelHB}                       & 87.7\cite{MadelHB}                   & 43.4\cite{MadelHB}                   & 40.76\cite{MadelHB}                  & -                  \\ 
     & Present       & 6.1877                                     & 86.39                                & 40.65                                & 40.71                                & 0.5893             \\ \hline\hline
     & Prev theory   & 4.460\cite{WrNe94}                         & 293\cite{Wright97}, 252\cite{LoMa07} & 159\cite{Wright97}, 129\cite{LoMa07} & 155\cite{Wright97}, 147\cite{LoMa07} & 0.61\cite{Wright97} \\ 
GaN  & Experimental  & 4.50597\cite{FrLo17}, 4.510\cite{NoZa10}   & -                                    & -                                    & -                                    & -                   \\ 
     & Present       & 4.4925                                     & 288.35                               & 152.98                               & 166.68                               & 0.5678              \\ \hline\hline
     & Prev theory   & 5.463\cite{RaMo15}, 5.322\cite{WaYe02}     & 142\cite{RaMo15}, 150.7\cite{WaYe03} & 61\cite{RaMo15}, 62.8\cite{WaYe03}    & 72\cite{RaMo15}, 76.3\cite{WaYe03} & 0.516\cite{WaYe03}  \\ 
GaP  & Experimental  & 5.439\cite{MaRo98}                         & 140\cite{MaRo98}                     & 62\cite{MaRo98}                      & 70\cite{MaRo98}                      & -                  \\ 
     & Present       & 5.4600                                     & 142.16                               & 60.47                                & 72.58                                & 0.5333             \\ \hline\hline
     & Prev theory   & 5.619\cite{SoSc98}, 5.75\cite{LoMa07}      & 125.6\cite{SoSc98}, 99\cite{LoMa07}  & 55.06\cite{SoSc98}, 41\cite{LoMa07}   & 60.56\cite{SoSc98}, 51\cite{LoMa07}   & 0.514\cite{SoSc98} \\ 
GaAs & Experimental  & 5.65325\cite{Blake82}                      & 113\cite{Blake82}, 120\cite{MaRo98}  & 57\cite{Blake82}, 53\cite{MaRo98}     & 60\cite{Blake82}, 60\cite{MaRo98}     & 0.55$\pm0.02$\cite{CoGe89} \\ 
     & Present       & 5.6859                                     & 116.81                               & 49.64                                & 59.76                                & 0.5288              \\ \hline\hline
     & Prev theory   & 5.981\cite{WaYe02}                         & 92.7\cite{WaYe03}                    & 38.7\cite{WaYe03}                    & 46.2\cite{WaYe03}                    & 0.530\cite{WaYe03}  \\ 
GaSb & Experimental  & 6.0959\cite{MadelHB}                       & 88.34\cite{MadelHB}                  & 40.23\cite{MadelHB}                  & 43.22\cite{MadelHB}                  & -                   \\ 
     & Present       & 6.1524                                     & 86.37                                & 36.55                                & 43.44                                & 0.5517              \\ \hline\hline
     & Prev theory   & 4.932\cite{WrNe95}                         & 187\cite{Wright97}, 159\cite{LoMa07} & 125\cite{Wright97}, 102\cite{LoMa07} & 86\cite{Wright97}, 78\cite{LoMa07}   & 0.80\cite{Wright97} \\ 
InN  & Experimental  & 4.98\cite{StCh93}, 5.01\cite{ScAs06}       & -                                    & -                                    & -                                    & -                   \\ 
     & Present       & 4.9908                                     & 185.20                               & 121.72                               & 91.49                                & 0.7474              \\ \hline\hline
     & Prev theory   & 5.899\cite{RaMo15}, 5.729\cite{WaYe02}     & 101\cite{RaMo15}, 109.5\cite{WaYe03} & 54\cite{RaMo15}, 55.7\cite{WaYe03}   & 48\cite{RaMo15}, 52.6\cite{WaYe03}   & 0.615\cite{WaYe03}  \\ 
InP  & Experimental  & 5.8687\cite{MadelHB}                       & 101.1\cite{MadelHB}                  & 56.1\cite{MadelHB}                   & 45.6\cite{MadelHB}                   & -                   \\ 
     & Present       & 5.9035                                     & 100.42                               & 53.72                                & 47.39                                & 0.6520       \\ \hline\hline
     & Prev theory   & 6.103\cite{RaMo15}, 5.921\cite{WaYe02}     & 86\cite{RaMo15}, 92.2\cite{WaYe03}, 72\cite{Lepkowski2008}   & 45\cite{RaMo15}, 46.5\cite{WaYe03}, 43\cite{Lepkowski2008} & 40\cite{RaMo15}, 44.4\cite{WaYe03}, 33\cite{Lepkowski2008} & 0.598\cite{WaYe03} \\ 
InAs & Experimental  & 6.0583\cite{MadelHB}                       & 83.29\cite{MadelHB}                  & 45.26\cite{MadelHB}                    & 39.59\cite{MadelHB}                    & -            \\ 
     & Present       & 6.1160                                     & 84.28                                & 44.72                                  & 39.66                                  & 0.6378       \\ \hline\hline
     & Prev theory   & 6.542\cite{RaMo15}, 6.346\cite{WaYe02}     & 67\cite{RaMo15}, 72.0\cite{WaYe03}   & 34\cite{RaMo15}, 35.4\cite{WaYe03}     & 30\cite{RaMo15}, 34.1\cite{WaYe03} & 0.603\cite{WaYe03} \\ 
InSb & Experimental  & 6.4794\cite{MadelHB}                       & 69.18\cite{MadelHB}                  & 37.88\cite{MadelHB}                    & 31.32\cite{MadelHB}                & 0.68\cite{CoGe91}  \\ 
     & Present       & 6.5625                                     & 64.97                                & 33.00                                  & 30.42                              & 0.6366             \\ \hline\hline
\end{tabular}
\caption{Elastic and structural properties of III-V compounds, where $C_{ij}$  are the 
second-order elastic constants, $a_{0}$ is the equilibrium lattice constant and 
$\zeta$ is Kleinman's internal strain parameter. All calculations have been 
performed with a cutoff energy of 600~eV, on a k-point grid density of $10\times10\times10$, and the stress
method is used to obtain the elastic constants.} \label{tab:SOEC}
\end{table*}
Table~\ref{tab:SOEC} presents a comprehensive comparison with experiment and previous theory
of lattice constants, SOECs, and Kleinman parameters for all considered materials.
The table reveals an abundance of both experimental and theoretical values of lattice and elastic constants
for all materials except for the metastable III-N compounds and highly toxic AlP, for which experimental elastic constants
are not available. For the Kleinman parameter, experimental values are rare, with measurements made only
on GaAs\cite{CoGe89} and InSb\cite{CoGe91}. The theoretical values presented are from DFT studies utilising different
approximations to the exchange correlation energy functional. Refs.~\onlinecite{WaYe02,WaYe03,WrNe94,WrNe95,Wright97,SoSc98}
use the local density approximation (LDA) to the exchange correlation functional. As is evident from the table,
in most cases LDA DFT accounts well for the elastic properties of solids; however, LDA is known to often overestimate
the binding in solids,\cite{RaMo15} resulting in smaller lattice and larger elastic constants. Indeed, we see from
Table~\ref{tab:SOEC} that whenever there is a significant disagreement between LDA 
elastic or lattice constants and those experimentally measured or here calculated, 
the LDA elastic constants tend to be larger. For the Al containing
compounds considered here this trend seems not to hold, with the elastic constants 
being often smaller than experiment, but nevertheless agreeing very closely. 
Refs~\onlinecite{LoMa07,Lepkowski2008} use the generalised gradient approximation (GGA) of Purdew,
Burke and Ernzerhof (PBE)\cite{PBE}. This functional tends to underestimate binding energies\cite{RaMo15,LoMa07},
and examining in particular InAs and GaAs, we see this trend borne out. From Ref.~\onlinecite{RaMo15}, we take
those structural and elastic properties calculated using HSE; these show good agreement with the HSE-DFT values
of the present study, and with experimental values. This good agreement with experiment demonstrates both the validity of the particular HSE-DFT
determined elastic constants presented here, and of the use of this method for the calculation of structural 
and elastic properties in general.

\begin{table*}[t] 
\centering  
\begin{tabular}{ccccccc}
\hline\hline
     & $C_{111}$(GPa)    & $C_{112}$(GPa)    & $C_{123}$(GPa)    & $C_{144}$(GPa)    & $C_{155}$(GPa)    & $C_{456}$(GPa)  \\ \hline
AlN  & -1125$\pm$3      & -1036$\pm$8       & -44$\pm$12        & 51$\pm$3          & -789$\pm$3        & -11.6$\pm$0.7    \\ 
AlP  & -595$\pm$4        & -428$\pm$4        & -103$\pm$6        & 14.9$\pm$0.9      & -243$\pm$1        & -33$\pm$1       \\ 
AlAs & -526$\pm$4        & -364$\pm$3        & -86$\pm$4         & 7.1$\pm$0.7       & -220$\pm$1        & -27$\pm$1       \\ 
AlSb & -416$\pm$3        & -268$\pm$2        & -77$\pm$3         & 6.4$\pm$0.5       & -156.9$\pm$0.6    & -21.4$\pm$0.7   \\ 
GaN  & -1277$\pm$8       & -976$\pm$4        & -252$\pm$9        & -46$\pm$1         & -647$\pm$2        & -49$\pm$1       \\ 
GaP  & -753$\pm$8        & -441$\pm$7        & -73$\pm$7         & -10$\pm$1         & -295$\pm$1        & -47$\pm$1       \\ 
GaAs & -612$\pm$5        & -351$\pm$4        & -86$\pm$5         & -15.2$\pm$0.9     & -264$\pm$1        & -33$\pm$1       \\ 
GaSb & -471$\pm$6        & -260$\pm$5        & -63$\pm$4         & 5$\pm$1           & -192$\pm$1        & -19.3$\pm$0.3   \\ 
InN  & -786$\pm$8        & -701$\pm$8        & -327$\pm$12       & 28$\pm$2          & -290$\pm$1        & 22$\pm$1        \\ 
InP  & -491$\pm$2.5      & -336$\pm$2        & -131$\pm$3.5      & -5.17$\pm$0.59    & -168.6$\pm$0.6    & -13.6$\pm$0.6   \\ 
InAs & -406$\pm$16       & -262$\pm$15       & -132$\pm$13       & -8.8$\pm$0.6      & -156$\pm$1        & -7.9$\pm$0.7    \\ 
InSb & -360$\pm$4        & -235$\pm$3        & -94$\pm$3         & -14$\pm$2         & -122$\pm$1        & -6.8$\pm$0.8    \\ \hline\hline
\end{tabular}
\caption{HSE-DFT calculated third-order elastic constants of selected III-V compounds. 
All calculations have been performed with a cutoff energy of 600~eV, and on a k-point grid density of $10\times10\times10$.
The elastic constants were extracted by fitting to the stresses, the given errors are fitting errors.} \label{tab:Elastic_Consts}
\end{table*}

 \begin{table*}[ht] 
\centering  
\begin{tabular}{ccccccc}
\hline\hline
        & $C_{111}$(GPa)                                & $C_{112}$(GPa)                                & $C_{123}$(GPa)                               & $C_{144}$(GPa)                              & $C_{155}$(GPa)                                & $C_{456}$(GPa)                              \\ \hline
   AlN  & $-$\emph{1070}\textsuperscript{a}               & $-$\emph{965}\textsuperscript{a}                & $-$\emph{61}\textsuperscript{a}                & \emph{57}\textsuperscript{a}                & $-$\emph{757}\textsuperscript{a}                & $-$\emph{9}\textsuperscript{a}      \\ 
   GaN  & $-$\emph{1213}\textsuperscript{a}               & $-$\emph{867}\textsuperscript{a}                & $-$\emph{253}\textsuperscript{a}               & $-$\emph{46}\textsuperscript{a}               & $-$\emph{606}\textsuperscript{a}                & $-$\emph{49}\textsuperscript{a}   \\ 
   GaP  & -676$\pm$52\textsuperscript{b,c}              & -499$\pm$25\textsuperscript{b,c}                & -82$\pm$56\textsuperscript{b,c}                & 75$\pm$47\textsuperscript{b,c}                & -332$\pm$23\textsuperscript{b,c}                & 199$\pm$66\textsuperscript{b,c}     \\ 
  GaAs  & $-$\emph{561}\textsuperscript{a},-618$\pm$9\textsuperscript{b,d,e} & $-$\emph{318}\textsuperscript{a},-389$\pm$4\textsuperscript{b,d,e} & $-$\emph{70}\textsuperscript{a},-48$\pm$11\textsuperscript{b,d,e} & $-$\emph{16}\textsuperscript{a},50$\pm$25\textsuperscript{b,d,e} & $-$\emph{242}\textsuperscript{a},-268$\pm$3\textsuperscript{b,d,e} & $-$\emph{22}\textsuperscript{a},-37$\pm$10\textsuperscript{b,d,e} \\ \
  GaSb  & -475$\pm$6\textsuperscript{f}                 & -308$\pm$2\textsuperscript{f}                 & -44$\pm$29\textsuperscript{f}                & 5$\pm$1\textsuperscript{f}                  & -216$\pm$13\textsuperscript{f}                & -25$\pm$15\textsuperscript{f}               \\ 
   InN  & $-$\emph{756}\textsuperscript{a}                       & $-$\emph{636}\textsuperscript{a}         & $-$\emph{310}\textsuperscript{a}               & \emph{13}\textsuperscript{a}                & $-$\emph{271}\textsuperscript{a}                & \emph{15}\textsuperscript{a}        \\ 
  InAs  & $-$\emph{404}\textsuperscript{g}                & $-$\emph{268}\textsuperscript{g}                & $-$\emph{121}\textsuperscript{a}                                         & $-$\emph{5}\textsuperscript{g}                & $-$\emph{138}\textsuperscript{g}                & $-$\emph{6}\textsuperscript{g}  \\ 
  InSb  & -338$\pm$30\textsuperscript{b,h}              & -242$\pm$17\textsuperscript{b,h}              & -79$\pm$14\textsuperscript{b,h}              & 13$\pm$7\textsuperscript{b,h}               & -131$\pm$7\textsuperscript{b,h}               & 0$\pm$3\textsuperscript{b,h}                \\ \hline\hline
\end{tabular}
\caption{Previous experimental and theoretical determinations of third-order elastic constants of GaAs and the cubic III-nitride materials. Theoretical values are italicised. 
a=Ref~\citenum{LoMa07}; b=Ref~\citenum{JoDu06}; c=Ref~\citenum{YoMi49} ; 
d= Ref~\citenum{DrBr67};   e=Ref~\citenum{McSkimAnd67}; f=Ref~\citenum{Raja_GaSb}; g=Ref~\citenum{Lepkowski2008}; h=Ref~\citenum{Raja_InSb}. } \label{tab:Prev_GaAs}
\end{table*}

\begin{figure}[t]
\centering\includegraphics[width=0.85\columnwidth]{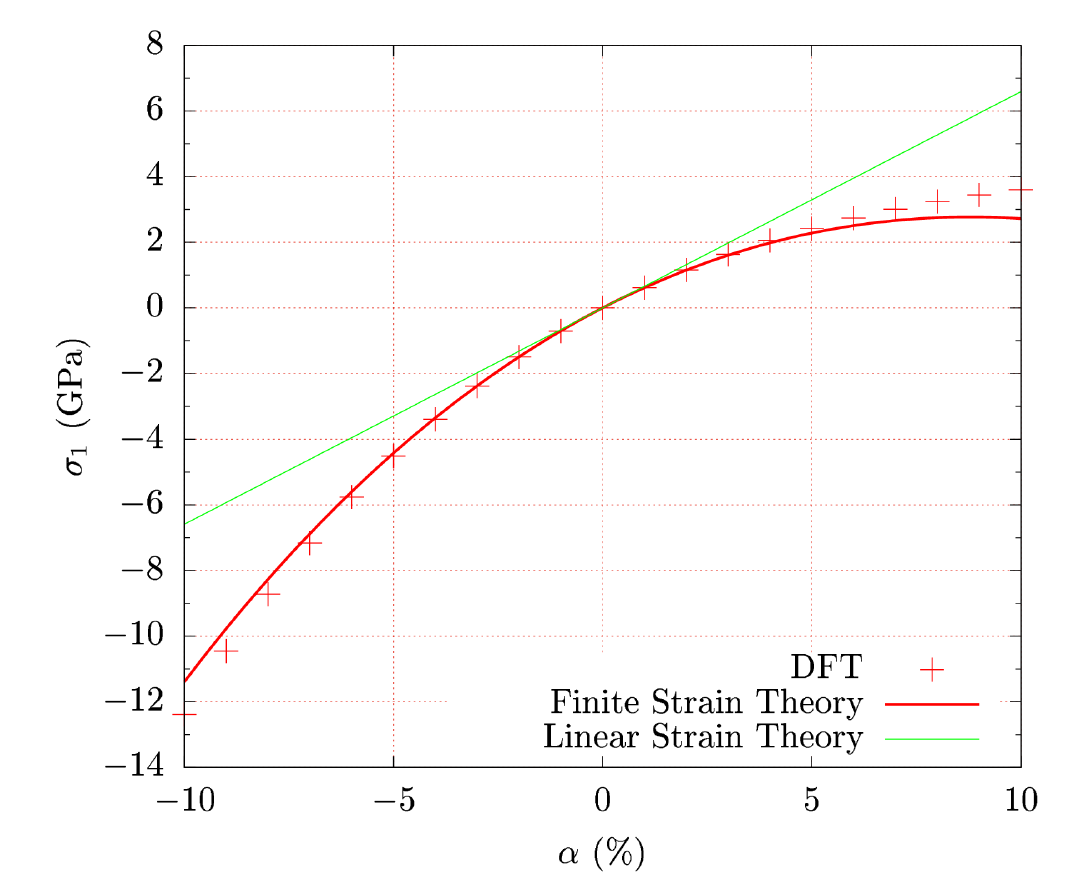}
\caption{Non-linear behaviour of $\sigma_{1}^{(4)}$ for InSb. Red crosses represent 
HSE-DFT data, whilst the red line shows the non-linear behaviour predicted from 
eqs.~(\ref{eq:t_AlphaBeta}) when $C_{ij}$ and $C_{ijk}$ are fitted on data sets out to
$\pm2$\% strain; the green line shows the behaviour at high strain predicted by linear 
infinitesimal strain theory. This is the stress that would be obtained in a biaxially 
strained system, strained equally along y and z axes. }
\label{fig:NolinInSbin}
\end{figure}

The TOECs, averaged over the independent determinations given in 
eqs.~(\ref{eq:t_eta}) and (\ref{eq:Lagrainge_Branches}), 
are gathered in Table~\ref{tab:Elastic_Consts}. The errors following the 
constants are the fitting errors.

In Table~\ref{tab:Prev_GaAs} experimental and theoretical values are provided 
for those materials for which they are available, with theoretical values
italicised. We find good agreement between the experimental measurements and our calculated values, 
taking into account that these measurements are performed often at room temperature ($T\approx300$~K) where
materials tend to be softer\cite{LoMa07} than at the $T=0$~K temperature at which DFT calculations
are made. With regard to literature theoretical calculations, for GaAs, there 
were several different works calculating TOECs.\cite{LoMa07,SoSc98,NiMa85,Lepkowski2008}
Here, we present only the most contemporary study, by \L{}opuszy\'{n}ski and Majewski.\cite{LoMa07} 
Overall, we find very good agreement between our results and those obtained via experiment or theory in the
literature. This serves as a validation for the extracted constants for which previous experimental or
theoretical values are not available.




With the TOECs and SOECs thus determined and validated against previous experimental
and theoretical values, we may use them to address the question of when third-order effects
become important in the materials under consideration. 
As a test case, we consider an InSb system that is strained in the $x-y$ plane and free to relax in the $z$
direction. In Fig.~\ref{fig:NolinInSbin},
the Cauchy stress, $\sigma$, in the $z$ direction,  of this system is
shown. This stress will be relevant to the pressure tuning of the Poisson ratio, 
and through this the pressure coefficient of the band-gap.
The figure plots the Cauchy stress, determined three different ways, against the strain. 
The stress obtained from DFT is given by the symbols, that obtained by linear strain theory is given
by the thin green line, and that obtained through third-order finite strain 
theory is given by the solid red line. Figure \ref{fig:NolinInSbin} shows 
clearly the increasing failure of the linear theory with increasing strain.
By $\pm$5\% strain, the linear theory suffers from errors in $\sigma$ of $26\%$ for $-5\%$ strain, 
and $45\%$ for $+5\%$ strain. This failing of the linear theory at these strains would 
introduce inaccuracies in the modelling of, for example, the elasticity of InSb/GaSb quantum wells\cite{QiWe93} and
QDs\cite{DeTa07} grown by the Stransky-Krastanov method, given that the lattice mismatch 
between InSb and GaSb is 6.3\%.

\begin{figure}[t]
\centering\includegraphics[width=0.85\columnwidth]{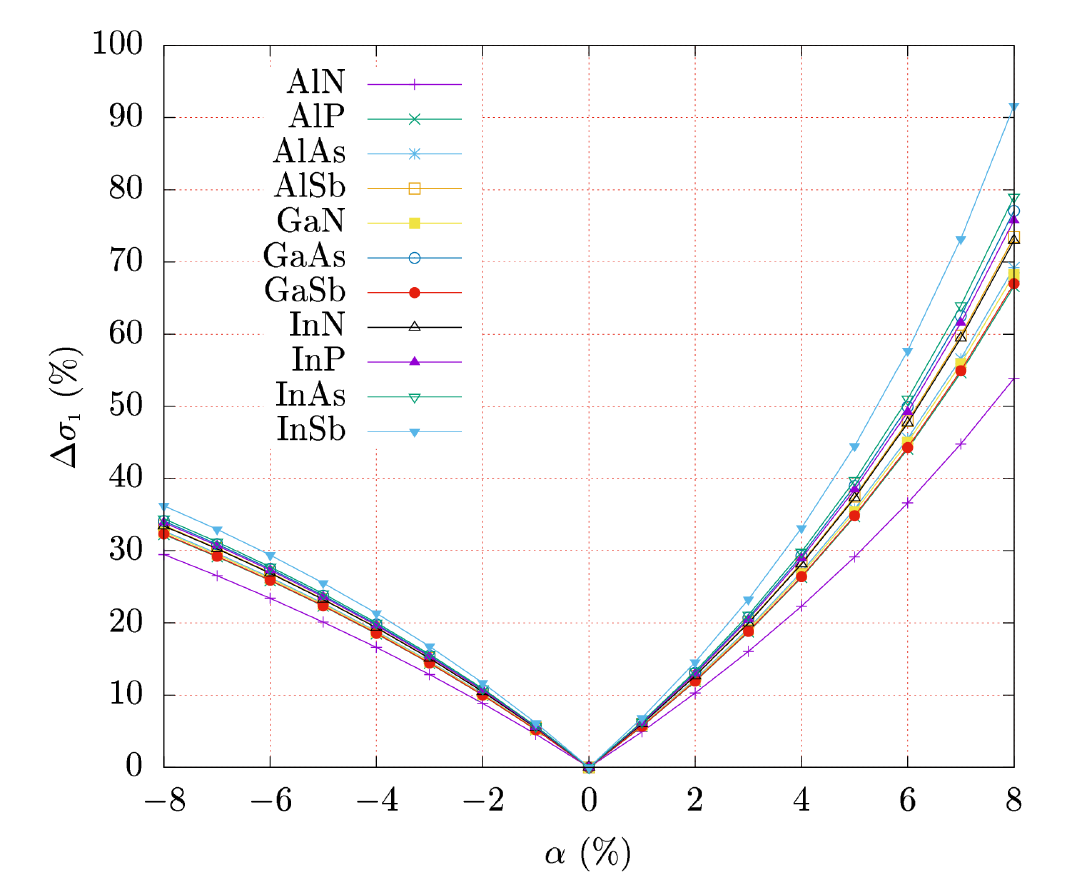}
\caption{Percentage error in $\sigma_{1}^{(4)}$, $\Delta\sigma = 
\frac{\sigma_{non-lin}-\sigma_{lin}}{\sigma_{non-lin}}$, where the error
is the difference between the stress as predicted using third-order finite strain
theory, $\sigma_{non-lin}$, and that predicted using a second-order infinitesimal
strain theory, $\sigma_{lin}$, as a function of increasing strain, $\alpha$.  }
\label{fig:NolinErr}
\end{figure}

Extending this analysis to the other materials, in Fig.\ref{fig:NolinErr}, the error in the out of plane stress
induced by a biaxial strain as calculated by the linear strain theory when compared with the non-linear theory
is plotted as a function of applied strain. From the figure we infer that once the strain in the system is 
greater than 2\%, the linear theory is no longer appropriate, with the errors 
in the stress being $\ge$10\% for all materials, except for AlN, which has a 9\% error 
in the calculated stress at $-$2\% strain. These large non-linearities in the
out of plane stress will manifest most noticeably in the pressure dependent behaviour 
of these materials in their respective heterostructures. Indeed, this has been 
already demonstrated by \L{}epkowski\cite{Lepkowski2008}.
From our results we may infer that the pressure tuning of strains in InSb/GaSb structures will be even more
markedly non-linear than that which has already been observed in InAs/GaAs and InN/GaN systems.

In the next section we turn to higher order effects in the internal strain.


\subsection{Internal strain tensor components} \label{Res:Inner}
The components of the internal strain tensor
are derived from eqs.~(\ref{eq:internal_Zeta}) and (\ref{eq:Lagrainge_Branches}).
These are given for the strain branches, $\boldsymbol{\eta}^{(1)}$, 
$\boldsymbol{\eta}^{(2)}$, and $\boldsymbol{\eta}^{(3)}$, in eq.~(\ref{eq:zeta_AlphaBeta}) below:
\begin{equation} \label{eq:zeta_AlphaBeta}
 \begin{aligned}
 \xi^{(1)}_{1} &= 
 \frac{1}{4}\left(A_{14} + 2A_{156}\right)\beta^{2} + A_{14}\beta , \\
 \xi^{(2)}_{1} &= A_{14}\beta +\frac{1}{2}A_{114}\alpha\beta , \\
 \xi^{(3)}_{1} &= \frac{1}{2}\left(A_{14}+A_{124}\right)\alpha\beta + A_{14}\beta .
 \end{aligned}
\end{equation}
The extracted non-zero components of the internal strain tensor are given in Table~\ref{tab:InternalStrain_Properties}. 
For $A_{14}$ the values from strain branches $\boldsymbol{\eta}^{(1)}$, $\boldsymbol{\eta}^{(2)}$ and $\boldsymbol{\eta}^{(3)}$, obtained
from fitting to eq.(~\ref{eq:zeta_AlphaBeta}) are averaged. Since
there is not the same abundance of equations from the relaxed atomic positions to describe the higher order internal 
strain tensor components as there are from the stresses for the elastic constants, the values for the 
different $A_{iJK}$ are set simply to those of the single independent determination of lowest error. 
For $A_{114}$ the only independent determination is that from $\boldsymbol{\eta}^{(2)}$; for $A_{156}$,
it is $\boldsymbol{\eta}^{(1)}$. For $A_{124}$ there are two independent determinations, but we include in the table only the value
from the uncomplicated $\boldsymbol{\eta}^{(3)}$ strain branch.

In terms of comparison with previous calculation or measurement, Table~\ref{tab:SOEC} reveals
very good agreement between our calculated first-order ISTC (the Kleinman parameter) and literature
values.
To the best of our knowledge, the only other first principles calculation of the components of the 
second-order internal strain tensor, in diamond or zincblende materials, are those obtained for C in Ref.~\onlinecite{Niel86}.
Strain derivatives of the Kleinman parameter are available for Si in Ref.~\citenum{NiMa85}, 
and for GaAs in Ref.~\citenum{MiPo06}. While these strain derivatives for the case of GaAs could be 
related to our data in Table~\ref{tab:InternalStrain_Properties}, we find that the obtained Kleinman
parameter of Ref.~\citenum{MiPo06}, 0.455,  disagrees significantly with our obtained value, 0.5288, with those from experiment, 
0.55$\pm$0.02,\cite{CousinsGaAsKlein} and with those from more recent theory 
0.514,\cite{SoSc98} 0.517;\cite{RaLe95} we do not therefore attempt explicit 
comparison.

\begin{table}[t] 
\centering  
\begin{tabular}{ccccc}
\hline\hline
      & $A_{14}$ (\AA{})         & $A_{114}$  (\AA{})      & $A_{124}$  (\AA{})       &   $A_{156}$ (\AA{}) \\ \hline
AlN   & -0.5888$\pm$0.0002  &    4.339$\pm$0.008 &    4.478$\pm$0.009  &    2.33$\pm$0.03 \\ 
AlP   & -0.7936$\pm$0.0003  &    4.01$\pm$0.01   &    5.32$\pm$0.01    &    1.81$\pm$0.06 \\ 
AlAs  & -0.8187$\pm$0.0003  &    3.959$\pm$0.009 &    5.385$\pm$0.007  &    1.95$\pm$0.06 \\ 
AlSb  & -0.9129$\pm$0.0003  &    3.95$\pm$0.01   &    5.53$\pm$0.01    &    1.72$\pm$0.06 \\ 
GaN   & -0.6394$\pm$0.0002  &    4.04$\pm$0.02   &    6.11$\pm$0.02    &    1.97$\pm$0.02 \\ 
GaP   & -0.7295$\pm$0.0002  &    3.417$\pm$0.008 &    5.65$\pm$0.01    &    1.98$\pm$0.04 \\ 
GaAs  & -0.7533$\pm$0.0005  &    3.584$\pm$0.009 &    5.55$\pm$0.01    &    2.38$\pm$0.08 \\ 
InN   & -0.9357$\pm$0.0002  &    5.12$\pm$0.05   &    6.61$\pm$0.04    &    1.23$\pm$0.03 \\ 
InP   & -0.9645$\pm$0.0004  &    3.86$\pm$0.03   &    6.72$\pm$0.03    &    1.44$\pm$0.07 \\ 
InAs  & -0.9777$\pm$0.0004  &    3.91$\pm$0.05   &    6.58$\pm$0.07    &    1.70$\pm$0.06 \\ 
InSb  & -1.0427$\pm$0.0009  &    3.19$\pm$0.25   &    6.61$\pm$0.07    &    1.8$\pm$0.1   \\ 
GaSb  & -0.8499$\pm$0.0002  &    3.48$\pm$0.02   &    5.38$\pm$0.01    &    2.20$\pm$0.03 \\ 
\hline\hline
\end{tabular}
\caption{Internal strain tensor components extracted from HSE-DFT data for Ga, In, and Al containing III-V compounds. 
Errors given are those associated with least squares fitting.} \label{tab:InternalStrain_Properties}
\end{table}

\section{Conclusion}
\label{sec:Summary}

In summary, second- and third- order elastic and first- and second- order internal 
strain tensor components were extracted from accurate HSE DFT calculations. 
The elastic constants and internal strain tensor components 
were extracted via stress-strain and position-strain
relations expressed within the formalism of finite strain, respectively.
This is the first determination of many of these constants. In particular,
the components of the second-order internal strain tensor extracted here have not 
before been measured or calculated.
Where previously determined, good agreement was obtained with experiment
and theory found in the literature. 
The results of convergence checks presented illustrate that far greater
care must be taken in the determination of third-order elastic constants (TOECs)
as compared to second-order elastic constants (SOECs),
with a high resolution of calculation required.
The use of the stress-strain equations for the calculation
of elastic constants was justified, and arguments from the literature,
formulated in the context of SOECs, were shown to have even more force in
the case of third-order elastic constants. 
The impact of non-linear strain effects was demonstrated in particular for the 
elasticity of InSb, and in general for other III-V materials systems, 
where it was found that third-order effects become significant for as little as 2\% strain.
Knowledge of the elastic constants and internal strain tensor components
presented here should therefore prove useful for the modelling of highly mismatched 
III-V heterostructures.

\section*{Acknowledgments}
This work was supported by Science Foundation Ireland (project
numbers 15/IA/3082 and 13/SIRG/2210) and by the European Union 7th
Framework Programme DEEPEN (grant agreement no.: 604416).

\bibliographystyle{apsrev4-1}
\bibliography{./Daniel_Tanner_Bibliography}

\begin{thebibliography}{80}%
\makeatletter
\providecommand \@ifxundefined [1]{%
 \@ifx{#1\undefined}
}%
\providecommand \@ifnum [1]{%
 \ifnum #1\expandafter \@firstoftwo
 \else \expandafter \@secondoftwo
 \fi
}%
\providecommand \@ifx [1]{%
 \ifx #1\expandafter \@firstoftwo
 \else \expandafter \@secondoftwo
 \fi
}%
\providecommand \natexlab [1]{#1}%
\providecommand \enquote  [1]{``#1''}%
\providecommand \bibnamefont  [1]{#1}%
\providecommand \bibfnamefont [1]{#1}%
\providecommand \citenamefont [1]{#1}%
\providecommand \href@noop [0]{\@secondoftwo}%
\providecommand \href [0]{\begingroup \@sanitize@url \@href}%
\providecommand \@href[1]{\@@startlink{#1}\@@href}%
\providecommand \@@href[1]{\endgroup#1\@@endlink}%
\providecommand \@sanitize@url [0]{\catcode `\\12\catcode `\$12\catcode
  `\&12\catcode `\#12\catcode `\^12\catcode `\_12\catcode `\%12\relax}%
\providecommand \@@startlink[1]{}%
\providecommand \@@endlink[0]{}%
\providecommand \url  [0]{\begingroup\@sanitize@url \@url }%
\providecommand \@url [1]{\endgroup\@href {#1}{\urlprefix }}%
\providecommand \urlprefix  [0]{URL }%
\providecommand \Eprint [0]{\href }%
\providecommand \doibase [0]{http://dx.doi.org/}%
\providecommand \selectlanguage [0]{\@gobble}%
\providecommand \bibinfo  [0]{\@secondoftwo}%
\providecommand \bibfield  [0]{\@secondoftwo}%
\providecommand \translation [1]{[#1]}%
\providecommand \BibitemOpen [0]{}%
\providecommand \bibitemStop [0]{}%
\providecommand \bibitemNoStop [0]{.\EOS\space}%
\providecommand \EOS [0]{\spacefactor3000\relax}%
\providecommand \BibitemShut  [1]{\csname bibitem#1\endcsname}%
\let\auto@bib@innerbib\@empty
\bibitem [{\citenamefont {O'Reilly}(1989)}]{Reilly89}%
  \BibitemOpen
  \bibfield  {author} {\bibinfo {author} {\bibfnamefont {E.~P.}\ \bibnamefont
  {O'Reilly}},\ }\href
  {http://iopscience.iop.org/article/10.1088/0268-1242/4/3/001/pdf} {\bibfield
  {journal} {\bibinfo  {journal} {Semiconductor Science and Technology}\
  }\textbf {\bibinfo {volume} {4}},\ \bibinfo {pages} {121} (\bibinfo {year}
  {1989})}\BibitemShut {NoStop}%
\bibitem [{\citenamefont {Nye}(1985)}]{Nye_Book}%
  \BibitemOpen
  \bibfield  {author} {\bibinfo {author} {\bibfnamefont {J.~F.}\ \bibnamefont
  {Nye}},\ }\href@noop {} {\emph {\bibinfo {title} {{Physical Properties of
  Crystals: Their Representation by Tensors and Matrices (2$^\textrm{nd}$
  ed.)}}}}\ (\bibinfo  {publisher} {Oxford University Press},\ \bibinfo {year}
  {1985})\BibitemShut {NoStop}%
\bibitem [{\citenamefont {Moram}\ and\ \citenamefont {Vickers}(2009)}]{MoVi09}%
  \BibitemOpen
  \bibfield  {author} {\bibinfo {author} {\bibfnamefont {M.~A.}\ \bibnamefont
  {Moram}}\ and\ \bibinfo {author} {\bibfnamefont {M.~E.}\ \bibnamefont
  {Vickers}},\ }\href {http://stacks.iop.org/0034-4885/72/i=3/a=036502}
  {\bibfield  {journal} {\bibinfo  {journal} {Reports on Progress in Physics}\
  }\textbf {\bibinfo {volume} {72}},\ \bibinfo {pages} {036502} (\bibinfo
  {year} {2009})}\BibitemShut {NoStop}%
\bibitem [{\citenamefont {Holec}\ \emph {et~al.}(2007)\citenamefont {Holec},
  \citenamefont {Costa}, \citenamefont {Kappers},\ and\ \citenamefont
  {Humphreys}}]{HoCo07}%
  \BibitemOpen
  \bibfield  {author} {\bibinfo {author} {\bibfnamefont {D.}~\bibnamefont
  {Holec}}, \bibinfo {author} {\bibfnamefont {P.}~\bibnamefont {Costa}},
  \bibinfo {author} {\bibfnamefont {M.}~\bibnamefont {Kappers}}, \ and\
  \bibinfo {author} {\bibfnamefont {C.}~\bibnamefont {Humphreys}},\ }\href
  {\doibase https://doi.org/10.1016/j.jcrysgro.2006.12.054} {\bibfield
  {journal} {\bibinfo  {journal} {Journal of Crystal Growth}\ }\textbf
  {\bibinfo {volume} {303}},\ \bibinfo {pages} {314 } (\bibinfo {year}
  {2007})},\ \bibinfo {note} {proceedings of the Fifth Workshop on Modeling in
  Crystal Growth}\BibitemShut {NoStop}%
\bibitem [{\citenamefont {Kittel}(2004)}]{Kittel_8th}%
  \BibitemOpen
  \bibfield  {author} {\bibinfo {author} {\bibfnamefont {C.}~\bibnamefont
  {Kittel}},\ }\href@noop {} {\emph {\bibinfo {title} {Introduction to Solid
  State Physics, (8$^\textrm{th}$ ed.)}}}\ (\bibinfo  {publisher} {Wiley},\
  \bibinfo {address} {New York},\ \bibinfo {year} {2004})\BibitemShut {NoStop}%
\bibitem [{\citenamefont {Brand}\ \emph {et~al.}(2015)\citenamefont {Brand},
  \citenamefont {Dufour}, \citenamefont {Heinrich},\ and\ \citenamefont
  {Josse}}]{Mems}%
  \BibitemOpen
  \bibfield  {author} {\bibinfo {author} {\bibfnamefont {O.}~\bibnamefont
  {Brand}}, \bibinfo {author} {\bibfnamefont {I.}~\bibnamefont {Dufour}},
  \bibinfo {author} {\bibfnamefont {S.~M.}\ \bibnamefont {Heinrich}}, \ and\
  \bibinfo {author} {\bibfnamefont {F.}~\bibnamefont {Josse}},\ }\href@noop {}
  {\emph {\bibinfo {title} {Resonant MEMS: Fundamentals, Implementation and
  Application}}}\ (\bibinfo  {publisher} {Wiley-VCH},\ \bibinfo {address}
  {Weinheim},\ \bibinfo {year} {2015})\BibitemShut {NoStop}%
\bibitem [{\citenamefont {Keating}(1966{\natexlab{a}})}]{Keating66}%
  \BibitemOpen
  \bibfield  {author} {\bibinfo {author} {\bibfnamefont {P.~N.}\ \bibnamefont
  {Keating}},\ }\href@noop {} {\bibfield  {journal} {\bibinfo  {journal}
  {Physical Review}\ }\textbf {\bibinfo {volume} {145}},\ \bibinfo {pages}
  {637} (\bibinfo {year} {1966}{\natexlab{a}})}\BibitemShut {NoStop}%
\bibitem [{\citenamefont {Born}\ and\ \citenamefont
  {Huang}(1954)}]{Born_dynamic}%
  \BibitemOpen
  \bibfield  {author} {\bibinfo {author} {\bibfnamefont {M.}~\bibnamefont
  {Born}}\ and\ \bibinfo {author} {\bibfnamefont {K.}~\bibnamefont {Huang}},\
  }\href@noop {} {\emph {\bibinfo {title} {{Dynamical theory of crystal
  lattices}}}},\ Oxford classic texts in the physical sciences\ (\bibinfo
  {publisher} {Clarendon Press},\ \bibinfo {address} {Oxford},\ \bibinfo {year}
  {1954})\BibitemShut {NoStop}%
\bibitem [{\citenamefont {Kleinman}(1962)}]{Klein62}%
  \BibitemOpen
  \bibfield  {author} {\bibinfo {author} {\bibfnamefont {L.}~\bibnamefont
  {Kleinman}},\ }\href@noop {} {\bibfield  {journal} {\bibinfo  {journal}
  {Physical Review}\ }\textbf {\bibinfo {volume} {128}},\ \bibinfo {pages}
  {2614} (\bibinfo {year} {1962})}\BibitemShut {NoStop}%
\bibitem [{\citenamefont {Cousins}(1978{\natexlab{a}})}]{CousinsInner}%
  \BibitemOpen
  \bibfield  {author} {\bibinfo {author} {\bibfnamefont {C.~S.~G.}\
  \bibnamefont {Cousins}},\ }\href@noop {} {\bibfield  {journal} {\bibinfo
  {journal} {Journal of Physics C: Solid State Physics}\ }\textbf {\bibinfo
  {volume} {11}},\ \bibinfo {pages} {4867} (\bibinfo {year}
  {1978}{\natexlab{a}})}\BibitemShut {NoStop}%
\bibitem [{\citenamefont {Birman}(1958)}]{Birman58}%
  \BibitemOpen
  \bibfield  {author} {\bibinfo {author} {\bibfnamefont {J.~L.}\ \bibnamefont
  {Birman}},\ }\href@noop {} {\bibfield  {journal} {\bibinfo  {journal}
  {Physical Review}\ }\textbf {\bibinfo {volume} {111}},\ \bibinfo {pages}
  {1510} (\bibinfo {year} {1958})}\BibitemShut {NoStop}%
\bibitem [{\citenamefont {Martin}(1972)}]{Martin_Piezo}%
  \BibitemOpen
  \bibfield  {author} {\bibinfo {author} {\bibfnamefont {R.~M.}\ \bibnamefont
  {Martin}},\ }\href {https://journals.aps.org/prb/pdf/10.1103/PhysRevB.5.1607}
  {\bibfield  {journal} {\bibinfo  {journal} {Physical Review B}\ }\textbf
  {\bibinfo {volume} {5}},\ \bibinfo {pages} {1607} (\bibinfo {year}
  {1972})}\BibitemShut {NoStop}%
\bibitem [{\citenamefont {Landau}\ and\ \citenamefont
  {Lifshitz}(1986)}]{LANDAU1986}%
  \BibitemOpen
  \bibfield  {author} {\bibinfo {author} {\bibfnamefont {L.}~\bibnamefont
  {Landau}}\ and\ \bibinfo {author} {\bibfnamefont {E.}~\bibnamefont
  {Lifshitz}},\ }\href@noop {} {\emph {\bibinfo {title} {Theory of Elasticity
  (Third Edition)}}}\ (\bibinfo  {publisher} {Butterworth-Heinemann},\ \bibinfo
  {year} {1986})\BibitemShut {NoStop}%
\bibitem [{\citenamefont {Pryor}\ \emph {et~al.}(1998)\citenamefont {Pryor},
  \citenamefont {Kim}, \citenamefont {Wang}, \citenamefont {Williamson},\ and\
  \citenamefont {Zunger}}]{PrKi98}%
  \BibitemOpen
  \bibfield  {author} {\bibinfo {author} {\bibfnamefont {C.}~\bibnamefont
  {Pryor}}, \bibinfo {author} {\bibfnamefont {J.}~\bibnamefont {Kim}}, \bibinfo
  {author} {\bibfnamefont {L.~W.}\ \bibnamefont {Wang}}, \bibinfo {author}
  {\bibfnamefont {A.~J.}\ \bibnamefont {Williamson}}, \ and\ \bibinfo {author}
  {\bibfnamefont {A.}~\bibnamefont {Zunger}},\ }\href@noop {} {\bibfield
  {journal} {\bibinfo  {journal} {Journal of Applied Physics}\ }\textbf
  {\bibinfo {volume} {83}},\ \bibinfo {pages} {2548} (\bibinfo {year}
  {1998})}\BibitemShut {NoStop}%
\bibitem [{\citenamefont {Frogley}\ \emph {et~al.}(2000)\citenamefont
  {Frogley}, \citenamefont {Downes},\ and\ \citenamefont {Dunstan}}]{FrDo00}%
  \BibitemOpen
  \bibfield  {author} {\bibinfo {author} {\bibfnamefont {M.~D.}\ \bibnamefont
  {Frogley}}, \bibinfo {author} {\bibfnamefont {J.~R.}\ \bibnamefont {Downes}},
  \ and\ \bibinfo {author} {\bibfnamefont {D.~J.}\ \bibnamefont {Dunstan}},\
  }\href {\doibase 10.1103/PhysRevB.62.13612} {\bibfield  {journal} {\bibinfo
  {journal} {Phys. Rev. B}\ }\textbf {\bibinfo {volume} {62}},\ \bibinfo
  {pages} {13612} (\bibinfo {year} {2000})}\BibitemShut {NoStop}%
\bibitem [{\citenamefont {Ellaway}\ and\ \citenamefont {Faux}(2002)}]{ElFa02}%
  \BibitemOpen
  \bibfield  {author} {\bibinfo {author} {\bibfnamefont {S.~W.}\ \bibnamefont
  {Ellaway}}\ and\ \bibinfo {author} {\bibfnamefont {D.~A.}\ \bibnamefont
  {Faux}},\ }\href@noop {} {\bibfield  {journal} {\bibinfo  {journal} {Journal
  of Applied Physics}\ }\textbf {\bibinfo {volume} {92}},\ \bibinfo {pages}
  {3027} (\bibinfo {year} {2002})}\BibitemShut {NoStop}%
\bibitem [{\citenamefont {Ma}\ \emph {et~al.}(2004)\citenamefont {Ma},
  \citenamefont {Wang}, \citenamefont {Su}, \citenamefont {Fang}, \citenamefont
  {Ding}, \citenamefont {Niu},\ and\ \citenamefont {Li}}]{MaWa04}%
  \BibitemOpen
  \bibfield  {author} {\bibinfo {author} {\bibfnamefont {B.~S.}\ \bibnamefont
  {Ma}}, \bibinfo {author} {\bibfnamefont {X.~D.}\ \bibnamefont {Wang}},
  \bibinfo {author} {\bibfnamefont {F.~H.}\ \bibnamefont {Su}}, \bibinfo
  {author} {\bibfnamefont {Z.~L.}\ \bibnamefont {Fang}}, \bibinfo {author}
  {\bibfnamefont {K.}~\bibnamefont {Ding}}, \bibinfo {author} {\bibfnamefont
  {Z.~C.}\ \bibnamefont {Niu}}, \ and\ \bibinfo {author} {\bibfnamefont
  {G.~H.}\ \bibnamefont {Li}},\ }\href {\doibase 10.1063/1.1635988} {\bibfield
  {journal} {\bibinfo  {journal} {Journal of Applied Physics}\ }\textbf
  {\bibinfo {volume} {95}},\ \bibinfo {pages} {933} (\bibinfo {year}
  {2004})}\BibitemShut {NoStop}%
\bibitem [{\citenamefont {\L{}epkowski}(2008)}]{Lepkowski2008}%
  \BibitemOpen
  \bibfield  {author} {\bibinfo {author} {\bibfnamefont {S.~P.}\ \bibnamefont
  {\L{}epkowski}},\ }\href {\doibase 10.1103/PhysRevB.78.153307} {\bibfield
  {journal} {\bibinfo  {journal} {Phys. Rev. B}\ }\textbf {\bibinfo {volume}
  {78}},\ \bibinfo {pages} {153307} (\bibinfo {year} {2008})}\BibitemShut
  {NoStop}%
\bibitem [{\citenamefont {\L{}opuszy\ifmmode~\acute{n}\else \'{n}\fi{}ski}\
  and\ \citenamefont {Majewski}(2007)}]{LoMa07}%
  \BibitemOpen
  \bibfield  {author} {\bibinfo {author} {\bibfnamefont {M.}~\bibnamefont
  {\L{}opuszy\ifmmode~\acute{n}\else \'{n}\fi{}ski}}\ and\ \bibinfo {author}
  {\bibfnamefont {J.~A.}\ \bibnamefont {Majewski}},\ }\href@noop {} {\bibfield
  {journal} {\bibinfo  {journal} {Physical Review B}\ }\textbf {\bibinfo
  {volume} {76}},\ \bibinfo {pages} {045202} (\bibinfo {year}
  {2007})}\BibitemShut {NoStop}%
\bibitem [{\citenamefont {\L{}epkowski}\ and\ \citenamefont
  {Majewski}(2004)}]{Lepkowski2004}%
  \BibitemOpen
  \bibfield  {author} {\bibinfo {author} {\bibfnamefont {S.}~\bibnamefont
  {\L{}epkowski}}\ and\ \bibinfo {author} {\bibfnamefont {J.}~\bibnamefont
  {Majewski}},\ }\href {\doibase https://doi.org/10.1016/j.ssc.2004.07.002}
  {\bibfield  {journal} {\bibinfo  {journal} {Solid State Communications}\
  }\textbf {\bibinfo {volume} {131}},\ \bibinfo {pages} {763 } (\bibinfo {year}
  {2004})}\BibitemShut {NoStop}%
\bibitem [{\citenamefont {\L{}epkowski}\ \emph {et~al.}(2005)\citenamefont
  {\L{}epkowski}, \citenamefont {Majewski},\ and\ \citenamefont
  {Jurczak}}]{LeMa05}%
  \BibitemOpen
  \bibfield  {author} {\bibinfo {author} {\bibfnamefont {S.~P.}\ \bibnamefont
  {\L{}epkowski}}, \bibinfo {author} {\bibfnamefont {J.~A.}\ \bibnamefont
  {Majewski}}, \ and\ \bibinfo {author} {\bibfnamefont {G.}~\bibnamefont
  {Jurczak}},\ }\href {\doibase 10.1103/PhysRevB.72.245201} {\bibfield
  {journal} {\bibinfo  {journal} {Phys. Rev. B}\ }\textbf {\bibinfo {volume}
  {72}},\ \bibinfo {pages} {245201} (\bibinfo {year} {2005})}\BibitemShut
  {NoStop}%
\bibitem [{\citenamefont {Shan}\ \emph {et~al.}(1998)\citenamefont {Shan},
  \citenamefont {Ager}, \citenamefont {Walukiewicz}, \citenamefont {Haller},
  \citenamefont {McCluskey}, \citenamefont {Johnson},\ and\ \citenamefont
  {Bour}}]{ShAg98}%
  \BibitemOpen
  \bibfield  {author} {\bibinfo {author} {\bibfnamefont {W.}~\bibnamefont
  {Shan}}, \bibinfo {author} {\bibfnamefont {J.~W.}\ \bibnamefont {Ager}},
  \bibinfo {author} {\bibfnamefont {W.}~\bibnamefont {Walukiewicz}}, \bibinfo
  {author} {\bibfnamefont {E.~E.}\ \bibnamefont {Haller}}, \bibinfo {author}
  {\bibfnamefont {M.~D.}\ \bibnamefont {McCluskey}}, \bibinfo {author}
  {\bibfnamefont {N.~M.}\ \bibnamefont {Johnson}}, \ and\ \bibinfo {author}
  {\bibfnamefont {D.~P.}\ \bibnamefont {Bour}},\ }\href {\doibase
  10.1103/PhysRevB.58.R10191} {\bibfield  {journal} {\bibinfo  {journal} {Phys.
  Rev. B}\ }\textbf {\bibinfo {volume} {58}},\ \bibinfo {pages} {R10191}
  (\bibinfo {year} {1998})}\BibitemShut {NoStop}%
\bibitem [{\citenamefont {Deguffroy}\ \emph {et~al.}(2007)\citenamefont
  {Deguffroy}, \citenamefont {Tasco}, \citenamefont {Baranov}, \citenamefont
  {Tournié}, \citenamefont {Satpati}, \citenamefont {Trampert}, \citenamefont
  {Dunaevskii}, \citenamefont {Titkov},\ and\ \citenamefont
  {Ramonda}}]{DeTa07}%
  \BibitemOpen
  \bibfield  {author} {\bibinfo {author} {\bibfnamefont {N.}~\bibnamefont
  {Deguffroy}}, \bibinfo {author} {\bibfnamefont {V.}~\bibnamefont {Tasco}},
  \bibinfo {author} {\bibfnamefont {A.~N.}\ \bibnamefont {Baranov}}, \bibinfo
  {author} {\bibfnamefont {E.}~\bibnamefont {Tournié}}, \bibinfo {author}
  {\bibfnamefont {B.}~\bibnamefont {Satpati}}, \bibinfo {author} {\bibfnamefont
  {A.}~\bibnamefont {Trampert}}, \bibinfo {author} {\bibfnamefont {M.~S.}\
  \bibnamefont {Dunaevskii}}, \bibinfo {author} {\bibfnamefont
  {A.}~\bibnamefont {Titkov}}, \ and\ \bibinfo {author} {\bibfnamefont
  {M.}~\bibnamefont {Ramonda}},\ }\href {\doibase 10.1063/1.2748872} {\bibfield
   {journal} {\bibinfo  {journal} {Journal of Applied Physics}\ }\textbf
  {\bibinfo {volume} {101}},\ \bibinfo {pages} {124309} (\bibinfo {year}
  {2007})}\BibitemShut {NoStop}%
\bibitem [{\citenamefont {Caro}\ \emph {et~al.}(2015)\citenamefont {Caro},
  \citenamefont {Schulz},\ and\ \citenamefont {O'Reilly}}]{CaSc15}%
  \BibitemOpen
  \bibfield  {author} {\bibinfo {author} {\bibfnamefont {M.~A.}\ \bibnamefont
  {Caro}}, \bibinfo {author} {\bibfnamefont {S.}~\bibnamefont {Schulz}}, \ and\
  \bibinfo {author} {\bibfnamefont {E.~P.}\ \bibnamefont {O'Reilly}},\
  }\href@noop {} {\bibfield  {journal} {\bibinfo  {journal} {Physical Review
  B}\ }\textbf {\bibinfo {volume} {91}},\ \bibinfo {pages} {075203} (\bibinfo
  {year} {2015})}\BibitemShut {NoStop}%
\bibitem [{\citenamefont {Migliorato}\ \emph {et~al.}(2006)\citenamefont
  {Migliorato}, \citenamefont {Powell}, \citenamefont {Cullis}, \citenamefont
  {Hammerschmidt},\ and\ \citenamefont {Srivastava}}]{MiPo06}%
  \BibitemOpen
  \bibfield  {author} {\bibinfo {author} {\bibfnamefont {M.~A.}\ \bibnamefont
  {Migliorato}}, \bibinfo {author} {\bibfnamefont {D.}~\bibnamefont {Powell}},
  \bibinfo {author} {\bibfnamefont {A.~G.}\ \bibnamefont {Cullis}}, \bibinfo
  {author} {\bibfnamefont {T.}~\bibnamefont {Hammerschmidt}}, \ and\ \bibinfo
  {author} {\bibfnamefont {G.~P.}\ \bibnamefont {Srivastava}},\ }\href
  {\doibase 10.1103/PhysRevB.74.245332} {\bibfield  {journal} {\bibinfo
  {journal} {Phys. Rev. B}\ }\textbf {\bibinfo {volume} {74}},\ \bibinfo
  {pages} {245332} (\bibinfo {year} {2006})}\BibitemShut {NoStop}%
\bibitem [{\citenamefont {Hiki}(1981)}]{HikiRev}%
  \BibitemOpen
  \bibfield  {author} {\bibinfo {author} {\bibfnamefont {Y.}~\bibnamefont
  {Hiki}},\ }\href {\doibase 10.1146/annurev.ms.11.080181.000411} {\bibfield
  {journal} {\bibinfo  {journal} {Annual Review of Materials Science}\ }\textbf
  {\bibinfo {volume} {11}},\ \bibinfo {pages} {51} (\bibinfo {year}
  {1981})}\BibitemShut {NoStop}%
\bibitem [{\citenamefont {Cousins}(2001)}]{CousinsThesis}%
  \BibitemOpen
  \bibfield  {author} {\bibinfo {author} {\bibfnamefont {C.~S.~G.}\
  \bibnamefont {Cousins}},\ }\emph {\bibinfo {title} {{Inner Elasticity and the
  Higher-Order Elasticity of some Diamond and Graphite Allotropes}}},\
  \href@noop {} {Ph.D. thesis},\ \bibinfo  {school} {University of Exeter}
  (\bibinfo {year} {2001})\BibitemShut {NoStop}%
\bibitem [{\citenamefont {Brugger}(1964)}]{Brug64}%
  \BibitemOpen
  \bibfield  {author} {\bibinfo {author} {\bibfnamefont {K.}~\bibnamefont
  {Brugger}},\ }\href@noop {} {\bibfield  {journal} {\bibinfo  {journal}
  {Physical Review}\ }\textbf {\bibinfo {volume} {133}},\ \bibinfo {pages}
  {A1611} (\bibinfo {year} {1964})}\BibitemShut {NoStop}%
\bibitem [{\citenamefont {Keating}(1966{\natexlab{b}})}]{KeatingThird}%
  \BibitemOpen
  \bibfield  {author} {\bibinfo {author} {\bibfnamefont {P.~N.}\ \bibnamefont
  {Keating}},\ }\href@noop {} {\bibfield  {journal} {\bibinfo  {journal}
  {Physical Review}\ }\textbf {\bibinfo {volume} {149}},\ \bibinfo {pages}
  {674} (\bibinfo {year} {1966}{\natexlab{b}})}\BibitemShut {NoStop}%
\bibitem [{\citenamefont {Birch}(1947)}]{Birch47}%
  \BibitemOpen
  \bibfield  {author} {\bibinfo {author} {\bibfnamefont {F.}~\bibnamefont
  {Birch}},\ }\href@noop {} {\bibfield  {journal} {\bibinfo  {journal}
  {Physical Review}\ }\textbf {\bibinfo {volume} {71}},\ \bibinfo {pages} {809}
  (\bibinfo {year} {1947})}\BibitemShut {NoStop}%
\bibitem [{\citenamefont {Wallace}(1970)}]{Wallace1970}%
  \BibitemOpen
  \bibfield  {author} {\bibinfo {author} {\bibfnamefont {D.~C.}\ \bibnamefont
  {Wallace}}\ }(\bibinfo  {publisher} {Academic Press},\ \bibinfo {year}
  {1970})\ pp.\ \bibinfo {pages} {301 -- 404}\BibitemShut {NoStop}%
\bibitem [{\citenamefont {Thurston}\ and\ \citenamefont
  {Brugger}(1964)}]{ThBr64}%
  \BibitemOpen
  \bibfield  {author} {\bibinfo {author} {\bibfnamefont {R.~N.}\ \bibnamefont
  {Thurston}}\ and\ \bibinfo {author} {\bibfnamefont {K.}~\bibnamefont
  {Brugger}},\ }\href {\doibase 10.1103/PhysRev.133.A1604} {\bibfield
  {journal} {\bibinfo  {journal} {Phys. Rev.}\ }\textbf {\bibinfo {volume}
  {133}},\ \bibinfo {pages} {A1604} (\bibinfo {year} {1964})}\BibitemShut
  {NoStop}%
\bibitem [{\citenamefont {Murnaghan}(1951)}]{Murnaghan_Book}%
  \BibitemOpen
  \bibfield  {author} {\bibinfo {author} {\bibfnamefont {F.~D.}\ \bibnamefont
  {Murnaghan}},\ }\href@noop {} {\emph {\bibinfo {title} {Finite Deformation Of
  An Elastic Solid}}}\ (\bibinfo  {publisher} {John Wiley \& Sons, Inc},\
  \bibinfo {year} {1951})\BibitemShut {NoStop}%
\bibitem [{\citenamefont {Murnaghan}(1937)}]{MurnaghanPaper}%
  \BibitemOpen
  \bibfield  {author} {\bibinfo {author} {\bibfnamefont {F.~D.}\ \bibnamefont
  {Murnaghan}},\ }\href {http://www.jstor.org/stable/2371405} {\bibfield
  {journal} {\bibinfo  {journal} {American Journal of Mathematics}\ }\textbf
  {\bibinfo {volume} {59}},\ \bibinfo {pages} {235} (\bibinfo {year}
  {1937})}\BibitemShut {NoStop}%
\bibitem [{\citenamefont {Johal}\ and\ \citenamefont {Dunstan}(2006)}]{JoDu06}%
  \BibitemOpen
  \bibfield  {author} {\bibinfo {author} {\bibfnamefont {A.~S.}\ \bibnamefont
  {Johal}}\ and\ \bibinfo {author} {\bibfnamefont {D.~J.}\ \bibnamefont
  {Dunstan}},\ }\href {\doibase 10.1103/PhysRevB.73.024106} {\bibfield
  {journal} {\bibinfo  {journal} {Phys. Rev. B}\ }\textbf {\bibinfo {volume}
  {73}},\ \bibinfo {pages} {024106} (\bibinfo {year} {2006})}\BibitemShut
  {NoStop}%
\bibitem [{\citenamefont {Cousins}\ \emph {et~al.}(1991)\citenamefont
  {Cousins}, \citenamefont {Gerward}, \citenamefont {Olsen}, \citenamefont
  {Sethi},\ and\ \citenamefont {Sheldon}}]{CoGe91}%
  \BibitemOpen
  \bibfield  {author} {\bibinfo {author} {\bibfnamefont {C.~S.~G.}\
  \bibnamefont {Cousins}}, \bibinfo {author} {\bibfnamefont {L.}~\bibnamefont
  {Gerward}}, \bibinfo {author} {\bibfnamefont {J.~S.}\ \bibnamefont {Olsen}},
  \bibinfo {author} {\bibfnamefont {S.~A.}\ \bibnamefont {Sethi}}, \ and\
  \bibinfo {author} {\bibfnamefont {B.~J.}\ \bibnamefont {Sheldon}},\ }\href
  {\doibase 10.1002/pssa.2211260115} {\bibfield  {journal} {\bibinfo  {journal}
  {physica status solidi (a)}\ }\textbf {\bibinfo {volume} {126}},\ \bibinfo
  {pages} {135} (\bibinfo {year} {1991})}\BibitemShut {NoStop}%
\bibitem [{\citenamefont {Lang}\ and\ \citenamefont {Gupta}(2011)}]{LaGu11}%
  \BibitemOpen
  \bibfield  {author} {\bibinfo {author} {\bibfnamefont {J.~M.}\ \bibnamefont
  {Lang}}\ and\ \bibinfo {author} {\bibfnamefont {Y.~M.}\ \bibnamefont
  {Gupta}},\ }\href {\doibase 10.1103/PhysRevLett.106.125502} {\bibfield
  {journal} {\bibinfo  {journal} {Phys. Rev. Lett.}\ }\textbf {\bibinfo
  {volume} {106}},\ \bibinfo {pages} {125502} (\bibinfo {year}
  {2011})}\BibitemShut {NoStop}%
\bibitem [{\citenamefont {Chandrasekaran}\ \emph {et~al.}()\citenamefont
  {Chandrasekaran}, \citenamefont {Mohanlal},\ and\ \citenamefont
  {Saravanan}}]{ChMo96}%
  \BibitemOpen
  \bibfield  {author} {\bibinfo {author} {\bibfnamefont {K.~S.}\ \bibnamefont
  {Chandrasekaran}}, \bibinfo {author} {\bibfnamefont {S.~K.}\ \bibnamefont
  {Mohanlal}}, \ and\ \bibinfo {author} {\bibfnamefont {R.}~\bibnamefont
  {Saravanan}},\ }\href {\doibase 10.1002/pssb.2221960102} {\bibfield
  {journal} {\bibinfo  {journal} {physica status solidi (b)}\ }\textbf
  {\bibinfo {volume} {196}},\ \bibinfo {pages} {3}}\BibitemShut {NoStop}%
\bibitem [{\citenamefont {Suzuki}\ \emph {et~al.}(1968)\citenamefont {Suzuki},
  \citenamefont {Granato},\ and\ \citenamefont {Thomas}}]{SuTe68}%
  \BibitemOpen
  \bibfield  {author} {\bibinfo {author} {\bibfnamefont {T.}~\bibnamefont
  {Suzuki}}, \bibinfo {author} {\bibfnamefont {A.~V.}\ \bibnamefont {Granato}},
  \ and\ \bibinfo {author} {\bibfnamefont {J.~F.}\ \bibnamefont {Thomas}},\
  }\href {\doibase 10.1103/PhysRev.175.766} {\bibfield  {journal} {\bibinfo
  {journal} {Phys. Rev.}\ }\textbf {\bibinfo {volume} {175}},\ \bibinfo {pages}
  {766} (\bibinfo {year} {1968})}\BibitemShut {NoStop}%
\bibitem [{\citenamefont {Nielsen}\ and\ \citenamefont
  {Martin}(1985)}]{NiMa85}%
  \BibitemOpen
  \bibfield  {author} {\bibinfo {author} {\bibfnamefont {O.~H.}\ \bibnamefont
  {Nielsen}}\ and\ \bibinfo {author} {\bibfnamefont {R.~M.}\ \bibnamefont
  {Martin}},\ }\href@noop {} {\bibfield  {journal} {\bibinfo  {journal}
  {Physical Review B}\ }\textbf {\bibinfo {volume} {32}},\ \bibinfo {pages}
  {3792} (\bibinfo {year} {1985})}\BibitemShut {NoStop}%
\bibitem [{\citenamefont {Hmiel}\ \emph {et~al.}(2016)\citenamefont {Hmiel},
  \citenamefont {Winey}, \citenamefont {Gupta},\ and\ \citenamefont
  {Desjarlais}}]{HmWi16}%
  \BibitemOpen
  \bibfield  {author} {\bibinfo {author} {\bibfnamefont {A.}~\bibnamefont
  {Hmiel}}, \bibinfo {author} {\bibfnamefont {J.~M.}\ \bibnamefont {Winey}},
  \bibinfo {author} {\bibfnamefont {Y.~M.}\ \bibnamefont {Gupta}}, \ and\
  \bibinfo {author} {\bibfnamefont {M.~P.}\ \bibnamefont {Desjarlais}},\ }\href
  {\doibase 10.1103/PhysRevB.93.174113} {\bibfield  {journal} {\bibinfo
  {journal} {Phys. Rev. B}\ }\textbf {\bibinfo {volume} {93}},\ \bibinfo
  {pages} {174113} (\bibinfo {year} {2016})}\BibitemShut {NoStop}%
\bibitem [{\citenamefont {Nielsen}(1986)}]{Niel86}%
  \BibitemOpen
  \bibfield  {author} {\bibinfo {author} {\bibfnamefont {O.~H.}\ \bibnamefont
  {Nielsen}},\ }\href@noop {} {\bibfield  {journal} {\bibinfo  {journal}
  {Physical Review B}\ }\textbf {\bibinfo {volume} {34}},\ \bibinfo {pages}
  {5808} (\bibinfo {year} {1986})}\BibitemShut {NoStop}%
\bibitem [{\citenamefont {S{\"o}rgel}\ and\ \citenamefont
  {Scherz}(1998)}]{SoSc98}%
  \BibitemOpen
  \bibfield  {author} {\bibinfo {author} {\bibfnamefont {J.}~\bibnamefont
  {S{\"o}rgel}}\ and\ \bibinfo {author} {\bibfnamefont {U.}~\bibnamefont
  {Scherz}},\ }\href@noop {} {\bibfield  {journal} {\bibinfo  {journal} {The
  European Physical Journal B-Condensed Matter and Complex Systems}\ }\textbf
  {\bibinfo {volume} {5}},\ \bibinfo {pages} {45} (\bibinfo {year}
  {1998})}\BibitemShut {NoStop}%
\bibitem [{\citenamefont {Heyd}\ \emph {et~al.}(2003)\citenamefont {Heyd},
  \citenamefont {Scuseria},\ and\ \citenamefont {Ernzerhof}}]{HeSc03}%
  \BibitemOpen
  \bibfield  {author} {\bibinfo {author} {\bibfnamefont {J.}~\bibnamefont
  {Heyd}}, \bibinfo {author} {\bibfnamefont {G.~E.}\ \bibnamefont {Scuseria}},
  \ and\ \bibinfo {author} {\bibfnamefont {M.}~\bibnamefont {Ernzerhof}},\
  }\href@noop {} {\bibfield  {journal} {\bibinfo  {journal} {The Journal of
  Chemical Physics}\ }\textbf {\bibinfo {volume} {118}},\ \bibinfo {pages}
  {8207} (\bibinfo {year} {2003})}\BibitemShut {NoStop}%
\bibitem [{\citenamefont {Voigt}(1928)}]{Voigt}%
  \BibitemOpen
  \bibfield  {author} {\bibinfo {author} {\bibfnamefont {W.}~\bibnamefont
  {Voigt}},\ }\href@noop {} {\emph {\bibinfo {title} {{Lehrbuch der
  Kristallphysik}}}}\ (\bibinfo  {publisher} {Teubner},\ \bibinfo {address}
  {Leipzig},\ \bibinfo {year} {1928})\BibitemShut {NoStop}%
\bibitem [{\citenamefont {Nielsen}\ and\ \citenamefont
  {Martin}(1983)}]{NiMa83}%
  \BibitemOpen
  \bibfield  {author} {\bibinfo {author} {\bibfnamefont {O.~H.}\ \bibnamefont
  {Nielsen}}\ and\ \bibinfo {author} {\bibfnamefont {R.~M.}\ \bibnamefont
  {Martin}},\ }\href {\doibase 10.1103/PhysRevLett.50.697} {\bibfield
  {journal} {\bibinfo  {journal} {Phys. Rev. Lett.}\ }\textbf {\bibinfo
  {volume} {50}},\ \bibinfo {pages} {697} (\bibinfo {year} {1983})}\BibitemShut
  {NoStop}%
\bibitem [{\citenamefont {Cousins}(1978{\natexlab{b}})}]{CousinsSymmetry}%
  \BibitemOpen
  \bibfield  {author} {\bibinfo {author} {\bibfnamefont {C.~S.~G.}\
  \bibnamefont {Cousins}},\ }\href@noop {} {\bibfield  {journal} {\bibinfo
  {journal} {Journal of Physics C: Solid State Physics}\ }\textbf {\bibinfo
  {volume} {11}},\ \bibinfo {pages} {4881} (\bibinfo {year}
  {1978}{\natexlab{b}})}\BibitemShut {NoStop}%
\bibitem [{\citenamefont {Caro}\ \emph {et~al.}(2013)\citenamefont {Caro},
  \citenamefont {Schulz},\ and\ \citenamefont {O’Reilly}}]{Miguel_Stress}%
  \BibitemOpen
  \bibfield  {author} {\bibinfo {author} {\bibfnamefont {M.~A.}\ \bibnamefont
  {Caro}}, \bibinfo {author} {\bibfnamefont {S.}~\bibnamefont {Schulz}}, \ and\
  \bibinfo {author} {\bibfnamefont {E.~P.}\ \bibnamefont {O’Reilly}},\
  }\href@noop {} {\bibfield  {journal} {\bibinfo  {journal} {Journal of
  Physics: Condensed Matter}\ }\textbf {\bibinfo {volume} {25}},\ \bibinfo
  {pages} {025803} (\bibinfo {year} {2013})}\BibitemShut {NoStop}%
\bibitem [{\citenamefont {tanner}(2017)}]{MyThesis}%
  \BibitemOpen
  \bibfield  {author} {\bibinfo {author} {\bibfnamefont {D.~S.~P.}\
  \bibnamefont {tanner}},\ }\emph {\bibinfo {title} {{A study of the elastic
  and electronic properties of III-nitride semiconductors}}},\ \href@noop {}
  {Ph.D. thesis},\ \bibinfo  {school} {University College Cork} (\bibinfo
  {year} {2017})\BibitemShut {NoStop}%
\bibitem [{\citenamefont {Kresse}\ and\ \citenamefont
  {Furthm\"uller}(1996)}]{KrFu96}%
  \BibitemOpen
  \bibfield  {author} {\bibinfo {author} {\bibfnamefont {G.}~\bibnamefont
  {Kresse}}\ and\ \bibinfo {author} {\bibfnamefont {J.}~\bibnamefont
  {Furthm\"uller}},\ }\href {\doibase 10.1103/PhysRevB.54.11169} {\bibfield
  {journal} {\bibinfo  {journal} {Phys. Rev. B}\ }\textbf {\bibinfo {volume}
  {54}},\ \bibinfo {pages} {11169} (\bibinfo {year} {1996})}\BibitemShut
  {NoStop}%
\bibitem [{\citenamefont {Henderson}\ \emph {et~al.}(2011)\citenamefont
  {Henderson}, \citenamefont {Paier},\ and\ \citenamefont
  {Scuseria}}]{HePa2011}%
  \BibitemOpen
  \bibfield  {author} {\bibinfo {author} {\bibfnamefont {T.~M.}\ \bibnamefont
  {Henderson}}, \bibinfo {author} {\bibfnamefont {J.}~\bibnamefont {Paier}}, \
  and\ \bibinfo {author} {\bibfnamefont {G.~E.}\ \bibnamefont {Scuseria}},\
  }\href@noop {} {\bibfield  {journal} {\bibinfo  {journal} {Physica Status
  Solidi (b)}\ }\textbf {\bibinfo {volume} {248}},\ \bibinfo {pages} {767}
  (\bibinfo {year} {2011})}\BibitemShut {NoStop}%
\bibitem [{\citenamefont {Paier}\ \emph {et~al.}(2006)\citenamefont {Paier},
  \citenamefont {Marsman}, \citenamefont {Hummer}, \citenamefont {Kresse},
  \citenamefont {Gerber},\ and\ \citenamefont {Ángyán}}]{PaMa06}%
  \BibitemOpen
  \bibfield  {author} {\bibinfo {author} {\bibfnamefont {J.}~\bibnamefont
  {Paier}}, \bibinfo {author} {\bibfnamefont {M.}~\bibnamefont {Marsman}},
  \bibinfo {author} {\bibfnamefont {K.}~\bibnamefont {Hummer}}, \bibinfo
  {author} {\bibfnamefont {G.}~\bibnamefont {Kresse}}, \bibinfo {author}
  {\bibfnamefont {I.~C.}\ \bibnamefont {Gerber}}, \ and\ \bibinfo {author}
  {\bibfnamefont {J.~G.}\ \bibnamefont {Ángyán}},\ }\href@noop {} {\bibfield
  {journal} {\bibinfo  {journal} {The Journal of Chemical Physics}\ }\textbf
  {\bibinfo {volume} {124}},\ \bibinfo {pages} {154709} (\bibinfo {year}
  {2006})}\BibitemShut {NoStop}%
\bibitem [{\citenamefont {R{\aa{}}sander}\ and\ \citenamefont
  {Moram}(2015)}]{RaMo15}%
  \BibitemOpen
  \bibfield  {author} {\bibinfo {author} {\bibfnamefont {M.}~\bibnamefont
  {R{\aa{}}sander}}\ and\ \bibinfo {author} {\bibfnamefont {M.~A.}\
  \bibnamefont {Moram}},\ }\href {\doibase 10.1063/1.4932334} {\bibfield
  {journal} {\bibinfo  {journal} {The Journal of Chemical Physics}\ }\textbf
  {\bibinfo {volume} {143}},\ \bibinfo {pages} {144104} (\bibinfo {year}
  {2015})}\BibitemShut {NoStop}%
\bibitem [{\citenamefont {Golesorkhtabar}\ \emph {et~al.}(2013)\citenamefont
  {Golesorkhtabar}, \citenamefont {Pavone}, \citenamefont {Spitaler},
  \citenamefont {Puschnig},\ and\ \citenamefont {Draxl}}]{ElaStic}%
  \BibitemOpen
  \bibfield  {author} {\bibinfo {author} {\bibfnamefont {R.}~\bibnamefont
  {Golesorkhtabar}}, \bibinfo {author} {\bibfnamefont {P.}~\bibnamefont
  {Pavone}}, \bibinfo {author} {\bibfnamefont {J.}~\bibnamefont {Spitaler}},
  \bibinfo {author} {\bibfnamefont {P.}~\bibnamefont {Puschnig}}, \ and\
  \bibinfo {author} {\bibfnamefont {C.}~\bibnamefont {Draxl}},\ }\href
  {\doibase https://doi.org/10.1016/j.cpc.2013.03.010} {\bibfield  {journal}
  {\bibinfo  {journal} {Computer Physics Communications}\ }\textbf {\bibinfo
  {volume} {184}},\ \bibinfo {pages} {1861 } (\bibinfo {year}
  {2013})}\BibitemShut {NoStop}%
\bibitem [{\citenamefont {Wright}\ and\ \citenamefont {Nelson}(1995)}]{WrNe95}%
  \BibitemOpen
  \bibfield  {author} {\bibinfo {author} {\bibfnamefont {A.~F.}\ \bibnamefont
  {Wright}}\ and\ \bibinfo {author} {\bibfnamefont {J.~S.}\ \bibnamefont
  {Nelson}},\ }\href@noop {} {\bibfield  {journal} {\bibinfo  {journal} {Phys.
  Rev. B}\ }\textbf {\bibinfo {volume} {51}},\ \bibinfo {pages} {7866}
  (\bibinfo {year} {1995})}\BibitemShut {NoStop}%
\bibitem [{\citenamefont {Wright}(1997)}]{Wright97}%
  \BibitemOpen
  \bibfield  {author} {\bibinfo {author} {\bibfnamefont {A.~F.}\ \bibnamefont
  {Wright}},\ }\href {\doibase 10.1063/1.366114} {\bibfield  {journal}
  {\bibinfo  {journal} {Journal of Applied Physics}\ }\textbf {\bibinfo
  {volume} {82}},\ \bibinfo {pages} {2833} (\bibinfo {year}
  {1997})}\BibitemShut {NoStop}%
\bibitem [{\citenamefont {As}(2010)}]{As2010}%
  \BibitemOpen
  \bibfield  {author} {\bibinfo {author} {\bibfnamefont {D.~J.}\ \bibnamefont
  {As}},\ }\href {\doibase 10.1117/12.846846} {\enquote {\bibinfo {title}
  {Recent developments on non-polar cubic group iii nitrides for optoelectronic
  applications},}\ } (\bibinfo {year} {2010})\BibitemShut {NoStop}%
\bibitem [{\citenamefont {Petrov}\ \emph {et~al.}(1992)\citenamefont {Petrov},
  \citenamefont {Mojab}, \citenamefont {Powell}, \citenamefont {Greene},
  \citenamefont {Hultman},\ and\ \citenamefont {Sundgren}}]{PeMo92}%
  \BibitemOpen
  \bibfield  {author} {\bibinfo {author} {\bibfnamefont {I.}~\bibnamefont
  {Petrov}}, \bibinfo {author} {\bibfnamefont {E.}~\bibnamefont {Mojab}},
  \bibinfo {author} {\bibfnamefont {R.~C.}\ \bibnamefont {Powell}}, \bibinfo
  {author} {\bibfnamefont {J.~E.}\ \bibnamefont {Greene}}, \bibinfo {author}
  {\bibfnamefont {L.}~\bibnamefont {Hultman}}, \ and\ \bibinfo {author}
  {\bibfnamefont {J.}~\bibnamefont {Sundgren}},\ }\href {\doibase
  10.1063/1.106943} {\bibfield  {journal} {\bibinfo  {journal} {Applied Physics
  Letters}\ }\textbf {\bibinfo {volume} {60}},\ \bibinfo {pages} {2491}
  (\bibinfo {year} {1992})}\BibitemShut {NoStop}%
\bibitem [{\citenamefont {Wang}\ and\ \citenamefont {Ye}(2002)}]{WaYe02}%
  \BibitemOpen
  \bibfield  {author} {\bibinfo {author} {\bibfnamefont {S.~Q.}\ \bibnamefont
  {Wang}}\ and\ \bibinfo {author} {\bibfnamefont {H.~Q.}\ \bibnamefont {Ye}},\
  }\href {\doibase 10.1103/PhysRevB.66.235111} {\bibfield  {journal} {\bibinfo
  {journal} {Phys. Rev. B}\ }\textbf {\bibinfo {volume} {66}},\ \bibinfo
  {pages} {235111} (\bibinfo {year} {2002})}\BibitemShut {NoStop}%
\bibitem [{\citenamefont {Singh}(1993)}]{SinghBook}%
  \BibitemOpen
  \bibinfo {editor} {\bibfnamefont {J.}~\bibnamefont {Singh}},\ ed.,\ \enquote
  {\bibinfo {title} {Physics of semiconductors and their heterostructures},}\
  in\ \href@noop {} {\emph {\bibinfo {booktitle} {Physics of Semiconductors and
  Their Heterostructures}}}\ (\bibinfo  {publisher} {McGraw-Hill, New York},\
  \bibinfo {address} {New York},\ \bibinfo {year} {1993})\BibitemShut {NoStop}%
\bibitem [{\citenamefont {Wang}\ and\ \citenamefont {Ye}()}]{WaYe03}%
  \BibitemOpen
  \bibfield  {author} {\bibinfo {author} {\bibfnamefont {S.~Q.}\ \bibnamefont
  {Wang}}\ and\ \bibinfo {author} {\bibfnamefont {H.~Q.}\ \bibnamefont {Ye}},\
  }\href {\doibase 10.1002/pssb.200301861} {\bibfield  {journal} {\bibinfo
  {journal} {physica status solidi (b)}\ }\textbf {\bibinfo {volume} {240}},\
  \bibinfo {pages} {45}}\BibitemShut {NoStop}%
\bibitem [{\citenamefont {Bessolov}\ \emph {et~al.}(1982)\citenamefont
  {Bessolov}, \citenamefont {Konnikov},\ and\ \citenamefont
  {Umanskii}}]{BeKo82}%
  \BibitemOpen
  \bibfield  {author} {\bibinfo {author} {\bibfnamefont {V.}~\bibnamefont
  {Bessolov}}, \bibinfo {author} {\bibfnamefont {S.}~\bibnamefont {Konnikov}},
  \ and\ \bibinfo {author} {\bibfnamefont {V.}~\bibnamefont {Umanskii}},\
  }\href@noop {} {\bibfield  {journal} {\bibinfo  {journal} {Fiz. Tverd. Tela}\
  }\textbf {\bibinfo {volume} {24}},\ \bibinfo {pages} {1528} (\bibinfo {year}
  {1982})}\BibitemShut {NoStop}%
\bibitem [{\citenamefont {Madelung}(2003)}]{MadelHB}%
  \BibitemOpen
  \bibinfo {editor} {\bibfnamefont {O.}~\bibnamefont {Madelung}},\ ed.,\
  \enquote {\bibinfo {title} {Semiconductors: Data handbook, 3$^{rd}$
  edition},}\ in\ \href@noop {} {\emph {\bibinfo {booktitle} {Semiconductors:
  Data Handbook, 3$^{rd}$ edition}}}\ (\bibinfo  {publisher} {Springer-Verlag
  GmbH, Heidelberg},\ \bibinfo {address} {Berlin, Heidelberg},\ \bibinfo {year}
  {2003})\BibitemShut {NoStop}%
\bibitem [{\citenamefont {Wright}\ and\ \citenamefont {Nelson}(1994)}]{WrNe94}%
  \BibitemOpen
  \bibfield  {author} {\bibinfo {author} {\bibfnamefont {A.~F.}\ \bibnamefont
  {Wright}}\ and\ \bibinfo {author} {\bibfnamefont {J.~S.}\ \bibnamefont
  {Nelson}},\ }\href {\doibase 10.1103/PhysRevB.50.2159} {\bibfield  {journal}
  {\bibinfo  {journal} {Phys. Rev. B}\ }\textbf {\bibinfo {volume} {50}},\
  \bibinfo {pages} {2159} (\bibinfo {year} {1994})}\BibitemShut {NoStop}%
\bibitem [{\citenamefont {Frentrup}\ \emph {et~al.}(2017)\citenamefont
  {Frentrup}, \citenamefont {Lee}, \citenamefont {Sahonta}, \citenamefont
  {Kappers}, \citenamefont {Massabuau}, \citenamefont {Gupta}, \citenamefont
  {Oliver}, \citenamefont {Humphreys},\ and\ \citenamefont {Wallis}}]{FrLo17}%
  \BibitemOpen
  \bibfield  {author} {\bibinfo {author} {\bibfnamefont {M.}~\bibnamefont
  {Frentrup}}, \bibinfo {author} {\bibfnamefont {L.~Y.}\ \bibnamefont {Lee}},
  \bibinfo {author} {\bibfnamefont {S.-L.}\ \bibnamefont {Sahonta}}, \bibinfo
  {author} {\bibfnamefont {M.~J.}\ \bibnamefont {Kappers}}, \bibinfo {author}
  {\bibfnamefont {F.}~\bibnamefont {Massabuau}}, \bibinfo {author}
  {\bibfnamefont {P.}~\bibnamefont {Gupta}}, \bibinfo {author} {\bibfnamefont
  {R.~A.}\ \bibnamefont {Oliver}}, \bibinfo {author} {\bibfnamefont {C.~J.}\
  \bibnamefont {Humphreys}}, \ and\ \bibinfo {author} {\bibfnamefont {D.~J.}\
  \bibnamefont {Wallis}},\ }\href
  {http://stacks.iop.org/0022-3727/50/i=43/a=433002} {\bibfield  {journal}
  {\bibinfo  {journal} {Journal of Physics D: Applied Physics}\ }\textbf
  {\bibinfo {volume} {50}},\ \bibinfo {pages} {433002} (\bibinfo {year}
  {2017})}\BibitemShut {NoStop}%
\bibitem [{\citenamefont {Novikov}\ \emph {et~al.}(2010)\citenamefont
  {Novikov}, \citenamefont {Zainal}, \citenamefont {Akimov}, \citenamefont
  {Staddon}, \citenamefont {Kent},\ and\ \citenamefont {Foxon}}]{NoZa10}%
  \BibitemOpen
  \bibfield  {author} {\bibinfo {author} {\bibfnamefont {S.~V.}\ \bibnamefont
  {Novikov}}, \bibinfo {author} {\bibfnamefont {N.}~\bibnamefont {Zainal}},
  \bibinfo {author} {\bibfnamefont {A.~V.}\ \bibnamefont {Akimov}}, \bibinfo
  {author} {\bibfnamefont {C.~R.}\ \bibnamefont {Staddon}}, \bibinfo {author}
  {\bibfnamefont {A.~J.}\ \bibnamefont {Kent}}, \ and\ \bibinfo {author}
  {\bibfnamefont {C.~T.}\ \bibnamefont {Foxon}},\ }\href {\doibase
  10.1116/1.3276426} {\bibfield  {journal} {\bibinfo  {journal} {Journal of
  Vacuum Science \& Technology B}\ }\textbf {\bibinfo {volume} {28}},\ \bibinfo
  {pages} {C3B1} (\bibinfo {year} {2010})}\BibitemShut {NoStop}%
\bibitem [{\citenamefont {Madelung}\ \emph {et~al.}(1998)\citenamefont
  {Madelung}, \citenamefont {R{\"o}ssler},\ and\ \citenamefont
  {Schulz}}]{MaRo98}%
  \BibitemOpen
  \bibinfo {editor} {\bibfnamefont {O.}~\bibnamefont {Madelung}}, \bibinfo
  {editor} {\bibfnamefont {U.}~\bibnamefont {R{\"o}ssler}}, \ and\ \bibinfo
  {editor} {\bibfnamefont {M.}~\bibnamefont {Schulz}},\ eds.,\ \enquote
  {\bibinfo {title} {Springer materials},}\ in\ \href@noop {} {\emph {\bibinfo
  {booktitle} {Springer Materials}}}\ (\bibinfo  {publisher} {Springer-Verlag
  GmbH, Heidelberg},\ \bibinfo {address} {Berlin, Heidelberg},\ \bibinfo {year}
  {1998})\ pp.\ \bibinfo {pages} {1--9}\BibitemShut {NoStop}%
\bibitem [{\citenamefont {Blakemore}(1982)}]{Blake82}%
  \BibitemOpen
  \bibfield  {author} {\bibinfo {author} {\bibfnamefont {J.~S.}\ \bibnamefont
  {Blakemore}},\ }\href {\doibase 10.1063/1.331665} {\bibfield  {journal}
  {\bibinfo  {journal} {Journal of Applied Physics}\ }\textbf {\bibinfo
  {volume} {53}},\ \bibinfo {pages} {R123} (\bibinfo {year}
  {1982})}\BibitemShut {NoStop}%
\bibitem [{\citenamefont {Cousins}\ \emph
  {et~al.}(1989{\natexlab{a}})\citenamefont {Cousins}, \citenamefont {Gerward},
  \citenamefont {Olsen}, \citenamefont {Selsmark}, \citenamefont {Sheldon},\
  and\ \citenamefont {Webster}}]{CoGe89}%
  \BibitemOpen
  \bibfield  {author} {\bibinfo {author} {\bibfnamefont {C.~S.~G.}\
  \bibnamefont {Cousins}}, \bibinfo {author} {\bibfnamefont {L.}~\bibnamefont
  {Gerward}}, \bibinfo {author} {\bibfnamefont {J.~S.}\ \bibnamefont {Olsen}},
  \bibinfo {author} {\bibfnamefont {B.}~\bibnamefont {Selsmark}}, \bibinfo
  {author} {\bibfnamefont {B.~J.}\ \bibnamefont {Sheldon}}, \ and\ \bibinfo
  {author} {\bibfnamefont {G.~E.}\ \bibnamefont {Webster}},\ }\href@noop {}
  {\bibfield  {journal} {\bibinfo  {journal} {Semiconductor Science and
  Technology}\ }\textbf {\bibinfo {volume} {4}},\ \bibinfo {pages} {333}
  (\bibinfo {year} {1989}{\natexlab{a}})}\BibitemShut {NoStop}%
\bibitem [{\citenamefont {Strite}\ \emph {et~al.}(1993)\citenamefont {Strite},
  \citenamefont {Chandrasekhar}, \citenamefont {Smith}, \citenamefont {Sariel},
  \citenamefont {Chen}, \citenamefont {Teraguchi},\ and\ \citenamefont
  {Morkoç}}]{StCh93}%
  \BibitemOpen
  \bibfield  {author} {\bibinfo {author} {\bibfnamefont {S.}~\bibnamefont
  {Strite}}, \bibinfo {author} {\bibfnamefont {D.}~\bibnamefont
  {Chandrasekhar}}, \bibinfo {author} {\bibfnamefont {D.~J.}\ \bibnamefont
  {Smith}}, \bibinfo {author} {\bibfnamefont {J.}~\bibnamefont {Sariel}},
  \bibinfo {author} {\bibfnamefont {H.}~\bibnamefont {Chen}}, \bibinfo {author}
  {\bibfnamefont {N.}~\bibnamefont {Teraguchi}}, \ and\ \bibinfo {author}
  {\bibfnamefont {H.}~\bibnamefont {Morkoç}},\ }\href {\doibase
  https://doi.org/10.1016/0022-0248(93)90605-V} {\bibfield  {journal} {\bibinfo
   {journal} {Journal of Crystal Growth}\ }\textbf {\bibinfo {volume} {127}},\
  \bibinfo {pages} {204 } (\bibinfo {year} {1993})}\BibitemShut {NoStop}%
\bibitem [{\citenamefont {Schörmann}\ \emph {et~al.}(2006)\citenamefont
  {Schörmann}, \citenamefont {As}, \citenamefont {Lischka}, \citenamefont
  {Schley}, \citenamefont {Goldhahn}, \citenamefont {Li}, \citenamefont
  {Löffler}, \citenamefont {Hetterich},\ and\ \citenamefont {Kalt}}]{ScAs06}%
  \BibitemOpen
  \bibfield  {author} {\bibinfo {author} {\bibfnamefont {J.}~\bibnamefont
  {Schörmann}}, \bibinfo {author} {\bibfnamefont {D.~J.}\ \bibnamefont {As}},
  \bibinfo {author} {\bibfnamefont {K.}~\bibnamefont {Lischka}}, \bibinfo
  {author} {\bibfnamefont {P.}~\bibnamefont {Schley}}, \bibinfo {author}
  {\bibfnamefont {R.}~\bibnamefont {Goldhahn}}, \bibinfo {author}
  {\bibfnamefont {S.~F.}\ \bibnamefont {Li}}, \bibinfo {author} {\bibfnamefont
  {W.}~\bibnamefont {Löffler}}, \bibinfo {author} {\bibfnamefont
  {M.}~\bibnamefont {Hetterich}}, \ and\ \bibinfo {author} {\bibfnamefont
  {H.}~\bibnamefont {Kalt}},\ }\href {\doibase 10.1063/1.2422913} {\bibfield
  {journal} {\bibinfo  {journal} {Applied Physics Letters}\ }\textbf {\bibinfo
  {volume} {89}},\ \bibinfo {pages} {261903} (\bibinfo {year}
  {2006})}\BibitemShut {NoStop}%
\bibitem [{\citenamefont {Perdew}\ \emph {et~al.}(1996)\citenamefont {Perdew},
  \citenamefont {Burke},\ and\ \citenamefont {Ernzerhof}}]{PBE}%
  \BibitemOpen
  \bibfield  {author} {\bibinfo {author} {\bibfnamefont {J.~P.}\ \bibnamefont
  {Perdew}}, \bibinfo {author} {\bibfnamefont {K.}~\bibnamefont {Burke}}, \
  and\ \bibinfo {author} {\bibfnamefont {M.}~\bibnamefont {Ernzerhof}},\
  }\href@noop {} {\bibfield  {journal} {\bibinfo  {journal} {Phys. Rev. Lett.}\
  }\textbf {\bibinfo {volume} {77}},\ \bibinfo {pages} {3865} (\bibinfo {year}
  {1996})}\BibitemShut {NoStop}%
\bibitem [{\citenamefont {Yoḡurtçu}\ \emph {et~al.}(1981)\citenamefont
  {Yoḡurtçu}, \citenamefont {Miller},\ and\ \citenamefont
  {Saunders}}]{YoMi49}%
  \BibitemOpen
  \bibfield  {author} {\bibinfo {author} {\bibfnamefont {Y.}~\bibnamefont
  {Yoḡurtçu}}, \bibinfo {author} {\bibfnamefont {A.}~\bibnamefont {Miller}},
  \ and\ \bibinfo {author} {\bibfnamefont {G.}~\bibnamefont {Saunders}},\
  }\href {\doibase https://doi.org/10.1016/0022-3697(81)90010-X} {\bibfield
  {journal} {\bibinfo  {journal} {Journal of Physics and Chemistry of Solids}\
  }\textbf {\bibinfo {volume} {42}},\ \bibinfo {pages} {49 } (\bibinfo {year}
  {1981})}\BibitemShut {NoStop}%
\bibitem [{\citenamefont {Drabble}\ and\ \citenamefont
  {Brammer}(1967)}]{DrBr67}%
  \BibitemOpen
  \bibfield  {author} {\bibinfo {author} {\bibfnamefont {J.~R.}\ \bibnamefont
  {Drabble}}\ and\ \bibinfo {author} {\bibfnamefont {A.~J.}\ \bibnamefont
  {Brammer}},\ }\href@noop {} {\bibfield  {journal} {\bibinfo  {journal}
  {Proceedings of the Physical Society}\ }\textbf {\bibinfo {volume} {91}},\
  \bibinfo {pages} {959} (\bibinfo {year} {1967})}\BibitemShut {NoStop}%
\bibitem [{\citenamefont {McSkimin}\ and\ \citenamefont
  {Jr.}(1967)}]{McSkimAnd67}%
  \BibitemOpen
  \bibfield  {author} {\bibinfo {author} {\bibfnamefont {H.~J.}\ \bibnamefont
  {McSkimin}}\ and\ \bibinfo {author} {\bibfnamefont {P.~A.}\ \bibnamefont
  {Jr.}},\ }\href@noop {} {\bibfield  {journal} {\bibinfo  {journal} {Journal
  of Applied Physics}\ }\textbf {\bibinfo {volume} {38}},\ \bibinfo {pages}
  {2610} (\bibinfo {year} {1967})}\BibitemShut {NoStop}%
\bibitem [{\citenamefont {Raja}\ and\ \citenamefont {Reddy}(1976)}]{Raja_GaSb}%
  \BibitemOpen
  \bibfield  {author} {\bibinfo {author} {\bibfnamefont {V.~S.}\ \bibnamefont
  {Raja}}\ and\ \bibinfo {author} {\bibfnamefont {P.~J.}\ \bibnamefont
  {Reddy}},\ }\href {\doibase https://doi.org/10.1016/0375-9601(76)90652-6}
  {\bibfield  {journal} {\bibinfo  {journal} {Physics Letters A}\ }\textbf
  {\bibinfo {volume} {56}},\ \bibinfo {pages} {215} (\bibinfo {year}
  {1976})}\BibitemShut {NoStop}%
\bibitem [{\citenamefont {Raja}\ and\ \citenamefont {Reddy}(1977)}]{Raja_InSb}%
  \BibitemOpen
  \bibfield  {author} {\bibinfo {author} {\bibfnamefont {V.}~\bibnamefont
  {Raja}}\ and\ \bibinfo {author} {\bibfnamefont {P.}~\bibnamefont {Reddy}},\
  }\href {\doibase https://doi.org/10.1016/0038-1098(77)91131-0} {\bibfield
  {journal} {\bibinfo  {journal} {Solid State Communications}\ }\textbf
  {\bibinfo {volume} {21}},\ \bibinfo {pages} {701 } (\bibinfo {year}
  {1977})}\BibitemShut {NoStop}%
\bibitem [{\citenamefont {Qian}\ and\ \citenamefont {Wessels}(1993)}]{QiWe93}%
  \BibitemOpen
  \bibfield  {author} {\bibinfo {author} {\bibfnamefont {L.~Q.}\ \bibnamefont
  {Qian}}\ and\ \bibinfo {author} {\bibfnamefont {B.~W.}\ \bibnamefont
  {Wessels}},\ }\href {\doibase 10.1063/1.109971} {\bibfield  {journal}
  {\bibinfo  {journal} {Applied Physics Letters}\ }\textbf {\bibinfo {volume}
  {63}},\ \bibinfo {pages} {628} (\bibinfo {year} {1993})}\BibitemShut
  {NoStop}%
\bibitem [{\citenamefont {Cousins}\ \emph
  {et~al.}(1989{\natexlab{b}})\citenamefont {Cousins}, \citenamefont {Gerward},
  \citenamefont {Olsen}, \citenamefont {Selsmark}, \citenamefont {Sheldon},\
  and\ \citenamefont {Webster}}]{CousinsGaAsKlein}%
  \BibitemOpen
  \bibfield  {author} {\bibinfo {author} {\bibfnamefont {C.~S.~G.}\
  \bibnamefont {Cousins}}, \bibinfo {author} {\bibfnamefont {L.}~\bibnamefont
  {Gerward}}, \bibinfo {author} {\bibfnamefont {J.~S.}\ \bibnamefont {Olsen}},
  \bibinfo {author} {\bibfnamefont {B.}~\bibnamefont {Selsmark}}, \bibinfo
  {author} {\bibfnamefont {B.~J.}\ \bibnamefont {Sheldon}}, \ and\ \bibinfo
  {author} {\bibfnamefont {G.~E.}\ \bibnamefont {Webster}},\ }\href@noop {}
  {\bibfield  {journal} {\bibinfo  {journal} {Semiconductor Science and
  Technology}\ }\textbf {\bibinfo {volume} {4}},\ \bibinfo {pages} {333}
  (\bibinfo {year} {1989}{\natexlab{b}})}\BibitemShut {NoStop}%
\bibitem [{\citenamefont {Raynolds}\ \emph {et~al.}(1995)\citenamefont
  {Raynolds}, \citenamefont {Levine},\ and\ \citenamefont {Wilkins}}]{RaLe95}%
  \BibitemOpen
  \bibfield  {author} {\bibinfo {author} {\bibfnamefont {J.~E.}\ \bibnamefont
  {Raynolds}}, \bibinfo {author} {\bibfnamefont {Z.~H.}\ \bibnamefont
  {Levine}}, \ and\ \bibinfo {author} {\bibfnamefont {J.~W.}\ \bibnamefont
  {Wilkins}},\ }\href {\doibase 10.1103/PhysRevB.51.10477} {\bibfield
  {journal} {\bibinfo  {journal} {Phys. Rev. B}\ }\textbf {\bibinfo {volume}
  {51}},\ \bibinfo {pages} {10477} (\bibinfo {year} {1995})}\BibitemShut
  {NoStop}%
\end{thebibliography}%

\end{document}